\documentclass[referee,pdflatex,sn-mathphys-num]{sn-jnl}
\makeatletter
\if@referee
\renewcommand\abstractfont{\reset@font\fontsize{9bp}{20bp}\selectfont\leftskip=24pt\rightskip=24pt\parfillskip=0pt plus 1fil}
\fi
\makeatother

\usepackage{amssymb}
\usepackage{amsmath}
\usepackage{amsfonts}
\usepackage{graphicx}
\usepackage{multirow}
\usepackage{makecell}
\usepackage{amsthm}
\usepackage{xcolor}%
\usepackage{textcomp}%
\usepackage{manyfoot}%
\usepackage{booktabs}%
\usepackage{listings}%
\usepackage{soul}
\usepackage{array}
\usepackage{tabularx}
\usepackage{threeparttable}
\usepackage{hyperref}
\renewcommand{\thetable}{\Roman{table}}

\begin{document}

\title[Article Title]{Bulk CMSX-4 Single-Crystal Builds via Laser-Directed Energy Deposition: Epitaxial Stability and Microstructure Evolution} 

\author*[1]{\fnm{Divya} \sur{Nalajala}}\email{divyan@iisc.ac.in}
\equalcont{These authors contributed equally to this work.}

\author*[1]{\fnm{Swapnil} \sur{Bhure}}\email{swapnilbhure@iisc.ac.in}
\equalcont{These authors contributed equally to this work.}

\author[2]{\fnm{R.} \sur{Ajithkannan}}\email{mm22d010@smail.iitm.ac.in}

\author[3]{\fnm{Shubhang} \sur{Galagali}}\email{shubgalagali.231mt049@nitk.edu.in}

\author[1]{\fnm{Abhik} \sur{Choudhury}}\email{abhiknc@iisc.ac.in}

\affil*[1]{\orgdiv{Department of Materials Engineering}, \orgname{Indian Institute of Science}, \orgaddress{\street{CV Raman Road}, \city{Bengaluru}, \postcode{560012}, \state{Karnataka}, \country{India}}}

\affil[2]{\orgdiv{Department of Metallurgical and Materials Engineering}, \orgname{Indian Institute of Technology Madras}, \orgaddress{\city{Chennai}, \postcode{600036}, \state{Tamil Nadu}, \country{India}}}

\affil[3]{\orgdiv{Department of Metallurgical and Materials Engineering}, \orgname{National Institute of Technology Karnataka}, \orgaddress{\city{Surathkal}, \postcode{575025}, \state{Karnataka}, \country{India}}}

\abstract{Single-crystal processing of Ni-based superalloys using laser-directed energy deposition (L-DED) is extremely challenging, given the interplay of a number of processing parameters. In this work, we develop strategies to maintain single-crystal growth in long cylindrical rod geometries by combining dynamic adjustment of laser power, informed by in-situ infrared thermal feedback, with a simulation-guided external cooling system. This combined approach ensures optimal heat extraction during additive manufacturing, maintaining near-steady-state conditions at the solidification interface and extending continuous epitaxial growth to nearly twice the build height achievable through process control alone. The developed strategy results in continuous single-crystal growth, as validated by electron backscatter diffraction (EBSD). The steady-state nature of solidification is revealed through the near-constant primary dendritic arm spacing.
Further, the heat-treatment is optimized by reducing the solutionizing time (by 47\%), resulting in faster eutectic dissolution. Comparison of the aged microstructures reveals a similar distribution of precipitates between the classical Bridgman- and L-DED-processed samples.}

\keywords{Laser-directed energy deposition (L-DED), CMSX-4 superalloy, Epitaxial growth, Stray grain, Heat treatment}

\maketitle

\section{Introduction}
Single-crystalline (SX) superalloys are crucial for aerospace and power generation industries because of their exceptional high-temperature performance \cite{Reed2006, duhl1980development}. CMSX-4 is a second-generation SX superalloy developed for turbine aerofoil applications, particularly for the high-pressure turbine section of aero-engines. It contains refractory elements such as Tungsten, Tantalum, and Rhenium, which contribute to creep strength by exhibiting relatively low atomic mobility at higher temperatures \cite{pollock2006nickel}. To meet increasingly demanding high-temperature performance, turbine blade designs have evolved over time, including the incorporation of intricate cooling channels 
thereby making the manufacturing process complex and multistep. To produce SX microstructures, traditional manufacturing employs investment casting with controlled directional solidification, using either seed-based or spiral grain-selector-based methods to achieve the desired crystallographic orientations \cite{Reed2006, pollock2006nickel, jiang2020research, matache2015investigation}. However, it produces a relatively coarser dendritic microstructure with a primary dendrite arm spacing of 200-500 $\mu$m, characterized by severe microsegregation of alloying elements and the presence of inter-dendritic eutectic phases, requiring long solutionizing heat-treatment time to achieve compositional and microstructural homogeneity \cite{Reed2006, pollock2006nickel, ramsperger2016microstructure}. These limitations, together with the increasing geometric complexity of the turbine blade components, motivate the development of alternate manufacturing approaches. 

Additive manufacturing (AM) offers a promising route for producing complex geometries in a single step, while the higher thermal gradients and solidification rates can promote the epitaxial growth \cite{Debroy2018, Mostafaei2023, ana2021}. However, maintaining the SX microstructure throughout the build in AM processing remains challenging. Unlike conventional directional solidification (DS), where the thermal gradient and solidification velocity are relatively controlled, AM involves repeated heating and cooling cycles as the material is deposited layer by layer. This dynamic solidification behavior makes it challenging to maintain epitaxial growth throughout the build.   

Several studies have investigated the mechanisms governing epitaxial growth under the dynamic thermal conditions of AM processing \cite{Liu2023, Mo2026review,shang2026review}. Single-crystal growth can be achieved by suppressing the nucleation and growth of stray grains ahead of the solidification interface \cite{david1989correlation, david1997welding}. The occurrence of these grains is controlled by the ratio of the thermal gradient (G) to the solidification velocity (V), where higher values of G/V can suppress the transition from columnar to equiaxed grains (CET) \cite{Gaumann1999, Kurz2001}. Studies have been conducted that relate processing conditions during additive manufacturing to solidification variables using process modeling techniques. These process modeling techniques account for the multiphysics problem involving fluid flow and heat transfer at the scale of the melt pool and reveal not only the shape of the solidification interface but also provide access to the solidification variables, namely the thermal gradient and the solidification velocity. Using these correlations between printing parameters and solidification variables, some authors have developed process-microstructure maps to predict the conditions that favor epitaxial growth \cite{Gaumann1999, Gaumann2001, vitek2004analysis, vitek2005effect, anderson2010origin, anderson2010stray, acharya2014coupled}. 

While printing with these parameters can result in epitaxial growth within a given layer, single-crystalline growth can only be achieved up to a certain height \cite{basak2017sle, Basak2016}. This is due to changes in solidification conditions from the substrate to the top of the melt pool during a single laser pass. The maximum height until which epitaxiality can be achieved is determined by the choice of laser power and scanning speed. Therefore, to maintain epitaxiality during bulk manufacturing, it is important to remelt the non-epitaxial regions of previous layers \cite{Buchbender2020, bhure2024superalloys, bhure2025acta}.  Insufficient heat input or larger build rates can leave misoriented grains from the previous layers and unmelted powder particles that provide potential sites for stray-grain nucleation \cite{liu2015effects, bhure2024toward, yang2025stray}. 

Solidification cracking is an additional complication in hard-to-weld materials, where the susceptibility to cracking must be considered when determining conditions for single-crystal growth. The challenge is that typical solutions to the solidification cracking problem involve preheating the substrate, thereby reducing thermal gradients and lowering thermal strain rates \cite{Liu2023}. However, this contradicts the conditions required for single-crystallinity, which require higher thermal gradients. Thus, a compromised strategy needs to be evolved that balances the height to which single-crystallinity can be achieved in a single layer, without introducing solidification cracks during printing.

While the processing maps are helpful for determining the initial processing parameters, it is important to recognize the dynamic nature of the process and the evolution of conditions at the solidification front during additive manufacturing. Excessive power densities can increase heat accumulation and reduce the thermal gradient as the build progresses, leading to a loss of single-crystallinity and topology \cite{bhure2025acta, nalajala2026Materialia}. The powder feed rate similarly affects conditions at the solidification interface and in the melt pool; excessive powder for a given laser power and scan speed may lead to incomplete melting of powder particles, which can act as potential nucleating sites for stray grains \cite{liang2017microstructural}. Finally, even when thermal conditions are held constant, the crystallographic orientation of the substrate relative to the scan direction affects the preferred dendrite growth direction \cite{CHEN2023118566, WANG2015283}.
In addition to thermal conditions, recent investigations suggest that microsegregation in the SX substrate can promote stray grains due to constitutional undercooling ahead of the solidification front \cite{wittenzellner2023, rausch2020}. 
Thus, identification of strategies for achieving single-crystal growth requires consideration of several factors that influence not only the microstructure but also the geometry of the build.

While these studies investigated epitaxial growth from the SX substrate, powder bed processes such as selective electron beam melting (SEBM) tailored scan strategies to promote grain selection, similar to the spiral grain selector used in conventional single-crystal processing \cite{Korner2018a, pistor2021single, gotterbarm2020fabrication}. Despite this knowledge, most reported laser-based AM studies demonstrated epitaxial growth in relatively small geometries where the thermal conditions remain largely the same. This motivates the exploration of laser-directed energy deposition (L-DED) as an alternative for fabricating large SX structures, owing to the flexibility in machine design and process control. Extending epitaxial growth to large structures presents an additional challenge in L-DED, as heat accumulation with increasing build height progressively compromises the local thermal conditions and build topology in addition to the local thermal gradients discussed above. 

In our previous work, we fabricated crack-free CMSX-4 cylindrical rods (10 mm in diameter and 50 mm in height) with near single-crystalline microstructures using the L-DED process \cite{bhure2024superalloys}. This work showed that the 800 W-1200 mm/min laser power-scan speed combination produces a flatter melt-pool, thereby promoting directional growth, and that preheating the substrate to 1000 \textdegree C before starting deposition, along with remelting laser scans after the first few deposited layers, helps to extend epitaxial growth without solidification cracking. 
This strategy helps reduce the fraction of stray grains, but achieving epitaxial growth throughout the build remains challenging, as increasing the number of preheat scans beyond an optimal value can reduce thermal gradients and promote stray grain formation.
Our subsequent works \cite{bhure2025acta, nalajala2026Materialia} further show that a lower-power regime promotes stray grain formation due to insufficient remelting of equiaxed grains in previously deposited layers, with unmelted powder particles serving as potential nucleation sites. In contrast, increasing the laser power ensures complete melting of the powder particles and remelting of the misoriented grains, thereby favoring epitaxial growth. However, excessive heat accumulation leads to loss of topology in the builds.  
These observations indicate that a constant laser power-scan speed combination cannot yield the desired epitaxial growth while maintaining the build's topology. 

Therefore, a strategy with dynamic adjustment of process parameters needs to be adopted for extending single-crystallinity. In this study, we present process strategies to extend epitaxial growth in CMSX-4 superalloy builds fabricated using the L-DED process in cylindrical geometries.
These strategies are developed using inputs from an infrared (IR) camera, which is correlated with changes in the microstructure. In addition, we design a process-simulation-guided external cooling setup to mitigate heat accumulation during deposition, thereby helping to maintain uniform solidification conditions. After deposition, we analyze the microstructure and crystallographic orientation, and perform heat-treatment optimization trials to homogenize microstructural and compositional heterogeneity that arises during solidification, as well as aging of the precipitate microstructure under classical two-step heat-treatment schedules. The results are compared with the precipitate microstructure in samples processed using directional solidification in a Bridgman furnace, followed by the established solutionizing and heat-treatment schedule for the CMSX-4 alloy.

\section{Experimental Details} 
\label{sec:experimental_details}
\begin{figure}
    \centering
    \includegraphics[width=0.6\linewidth]{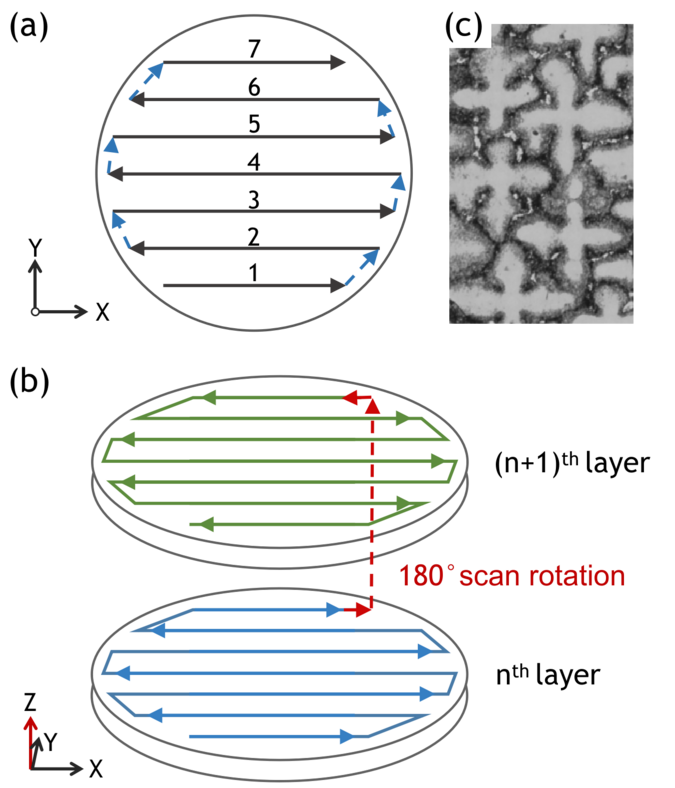}
    \caption{Schematic diagrams showing (a) bidirectional scan strategy within a layer, and (b) 180\textdegree scan rotation after every layer. (c) Optical macrograph of the top surface of the substrate showing the orientation of the secondary dendrite arm relative to the scan direction.
    } \label{fig:scan_strategy_schematic}
\end{figure}

CMSX-4 cylindrical rods of 10 mm in diameter and 100 mm in height are printed on a single-crystalline CMSX-4 substrate of dimensions 11.5 mm $\times$ 11.5 mm $\times$ 4 mm, using gas-atomized CMSX-4 powder on a FormAlloy L-DED machine. While L-DED involves several process variables, including laser power, scan speed, powder feed rate, layer thickness, hatch spacing, laser spot size, and scan strategy, we vary only laser power and scan speed in this study, holding all other parameters constant and optimizing layer height to maintain a constant working distance of 9 mm. We use Mastercam 2024 software to generate G-codes with a bidirectional scan strategy in a layer and 180\textdegree scan rotation after every layer, as depicted in the schematic Fig. \ref{fig:scan_strategy_schematic}(a)-(b). 
Before printing, the substrate is chemically macro-etched (as described below) and oriented in such a way that the scanning direction is parallel to secondary dendrite arms, as shown in Fig. \ref{fig:scan_strategy_schematic} (c).
To capture thermal data during printing, we use an Optris PI05M infrared (IR) camera operating in the spectral range of 500-540 nm, with a spatial resolution of 382 $\times$ 288 pixels at a frame rate of 10 Hz. We extract thermal gradients and melt-pool depths from the thermal data, plot them as a function of build height, and correlate them with the corresponding microstructures.

After printing, we cut the cylinders parallel to the scan direction using an Electrical Discharge Machining (EDM) system. We then macro-etch the cut samples by soaking them in a mixture of 80 g of ferric chloride ($\mathrm{FeCl_3}$), 40 ml of hydrochloric acid (HCl), 1 ml of nitric acid ($\mathrm{HNO_3}$), and 9 ml of water for about 20 min to reveal the grain structure and capture full-length images of the etched samples using a smartphone camera. For metallographic analysis, we polish the samples with emery papers graded from 220 to 5000 grit, followed by cloth polishing with a 1-micron alumina solution, and chemically etch them with a mixture of 1 ml of nitric acid ($\mathrm{HNO_3}$), 1.5 ml of Hydrochloric acid (HCl), and 1 ml of acetic acid to reveal the microstructure. Images of the resulting microstructure are captured using a Leica Optical microscope. Subsequently, the samples are re-polished with intermediate-grit emery papers to remove the etched surface, then polished in a vibratory polishing machine with 0.05-micron colloidal silica and distilled water (1:8 ratio) for about 6 hours to enable crystallographic orientation analysis using the Electron Backscatter Diffraction (EBSD) technique. We use the TSL OIM 8 software for the EBSD data analysis. Next, we capture the segregation behavior of various alloying elements using electron microprobe analysis (EPMA).
Furthermore, we conduct heat-treatment studies using a vertical resistive-heating furnace equipped with MoSi$_2$ heating elements and an integrated turbo-molecular pump, with heat treatment performed under a vacuum of ($2.5 \times 10^{-5}$ mbar), followed by quenching in a controlled environment by purging argon gas at 2 bar. Microstructural characterization of the heat-treated samples is performed using a TESCAN field-emission scanning electron microscope. The $\gamma\prime$ precipitate size is determined by measuring the area of approximately 280-300 precipitates from representative micrographs. The equivalent circular diameter is calculated by equating the measured precipitate area to the area of a circle, and the mean size and standard deviation are calculated from the measured population.    
\section{Dynamic parameter control for epitaxial stability}
\begin{figure*}[!htb]%
    \centering
    \includegraphics[width=0.99\linewidth]{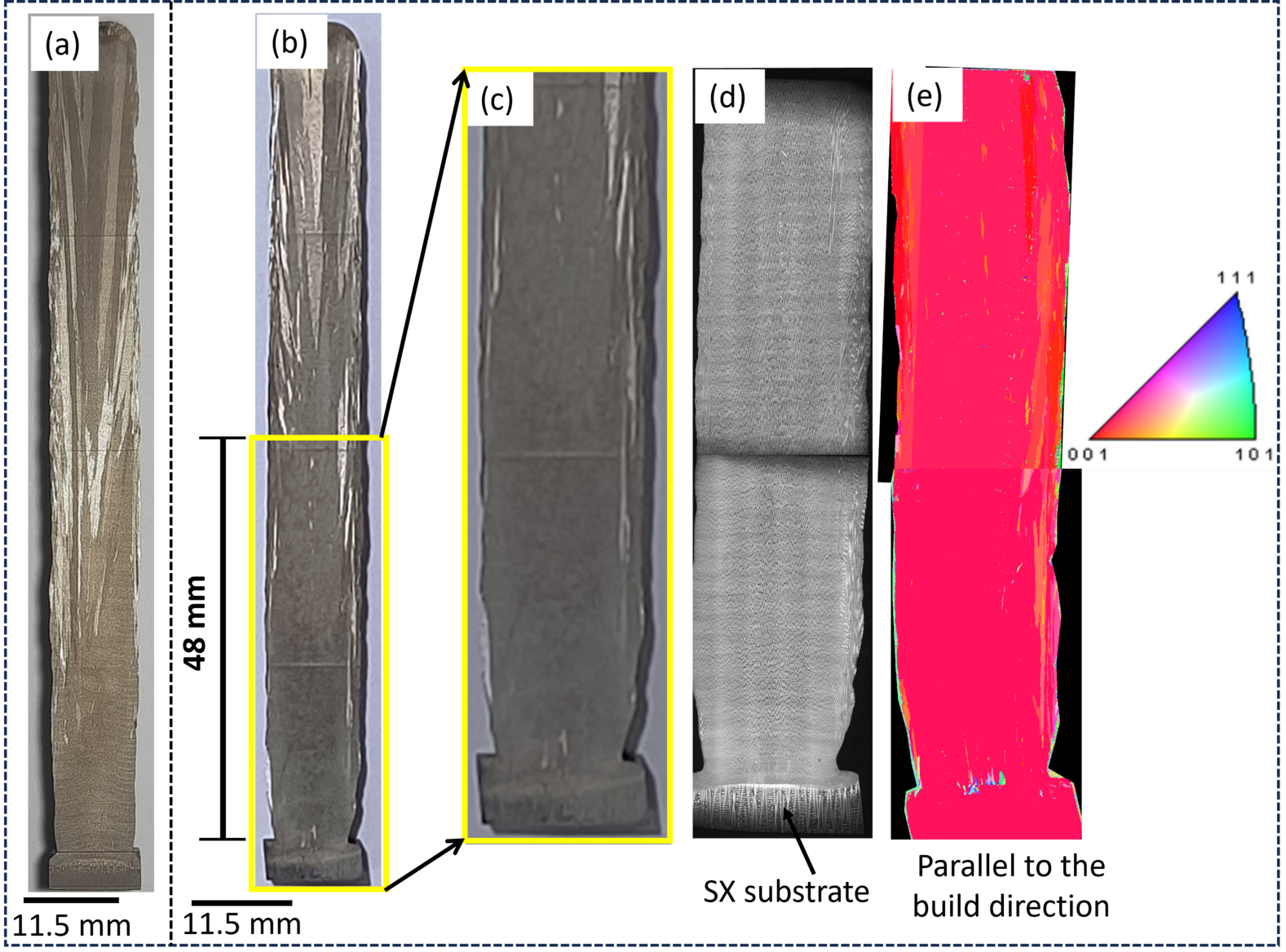}
    \caption{(a) Full-length macro-etched image revealing the grain structure of the rod printed with abrupt variation in the laser power; Constant scan speed of 1400 mm/min, varied laser power for first 5 layers - 1200 W, next 5 layers - 1000 W, remaining height with 800 W and (b), another rod with very gradual variation in the laser power; 1200 W-1000 W reduction over 15 mm (9.52 W/layer), constant 1000 W up to 40 mm, remaining build height with 1000 W-850 W reduction in power (1.76 W/layer) and (c) shows the enlarged image until about 48 mm build height and (d),(e) shows the corresponding optical micrograph and colored inverse pole figure (IPF) map parallel to the build direction, respectively. Colored IPF maps are post-processed with a 10\textdegree grain average to remove noise from stitching of tens of scans.
    } \label{fig:dynamic process control}
\end{figure*}

\begin{figure}[!htb]%
    \centering
    \includegraphics[width=0.6\linewidth]{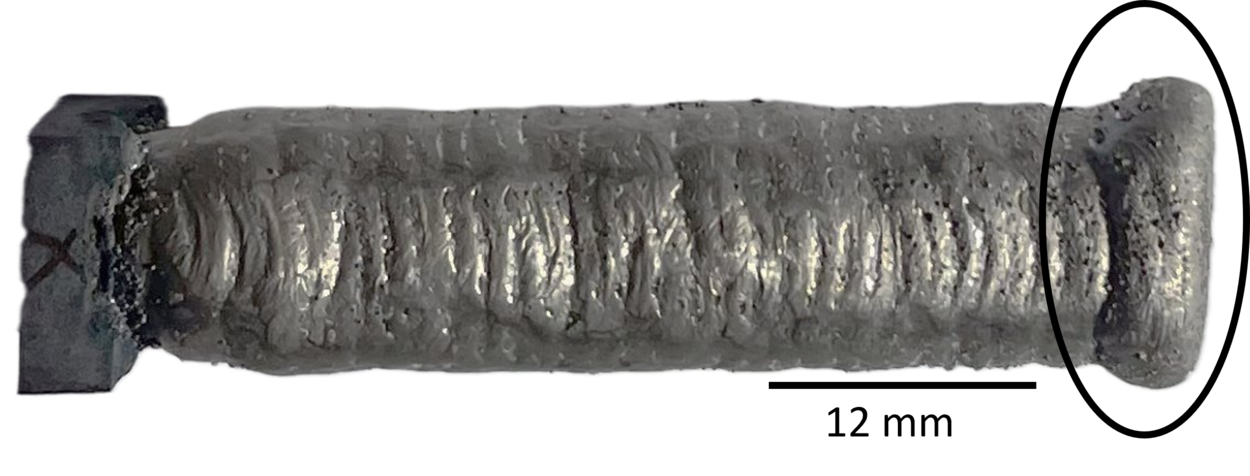}
    \caption{Photograph of a CMSX-4 rod printed with a scan speed of 1400 mm/min and a reduction in laser power from 1200 W to 1000 W over the first 15 mm and maintaining 1000 W constant thereafter, showing liquid overflow at the top portion due to excessive heat accumulation.
    } \label{fig:liquid_overflow}
\end{figure}

As shown in our previous work \cite{bhure2024superalloys}, an 800 W-1200 mm/min combination, together with remelting laser scans after the first few deposited layers, helps reduce the fraction of stray grains. However, isolated stray grains persist in the central region up to approximately 9 mm height from the bottom, resulting in non-uniform epitaxial growth. We therefore begin by increasing the laser power to 1200 W \cite{bhure2025acta} to promote clean epitaxial growth from the bottom. Since thermal gradients scale with scan speed, we fix the scan speed at 1400 mm/min (the maximum achievable on our machine) and keep the laser power as the only control variable, reducing it as we build to control heat accumulation and, in turn, increase the thermal gradients.  
We conduct an experiment with a laser power of 1200 W for the first five layers, then 1000 W for the next five layers, and 800 W for the remainder of the build. The corresponding macro-etched image shown in Fig. \ref{fig:dynamic process control}(a) reveals a single color up to a build height of approximately 20 mm, indicating successful epitaxial growth (the color contrast in the image represents grains misoriented relative to the substrate).
This confirms that the higher laser power–scan speed combination results in cleaner epitaxial growth near the base. However, as the build progresses, stray grains form, indicating that the abrupt reduction in laser power to 800 W changes the local solidification conditions, thereby interrupting sustained epitaxial growth. 
To mitigate the effects of abrupt changes in laser power, we vary it linearly with build height. We conduct multiple experiments to extend the epitaxial growth regime by systematically adjusting the laser power gradient per layer, while maintaining an initial deposition power of 1200 W. A more gradual reduction in laser power promotes sustained epitaxial growth to greater build heights without compromising the build topology. One optimized parameter set is discussed here: a reduction from 1200 W to 1000 W over the first 15 mm (corresponding to a gradient of approximately 9.52 W/layer), followed by deposition at 1000 W up to a build height of 40 mm (the maximum height achievable without liquid overflow due to excessive heat accumulation, as shown in Fig. \ref{fig:liquid_overflow}).
The remaining 60 mm of the build is deposited with a gradual reduction in laser power to 850 W at a gradient of 1.76 W/layer (to improve the thermal gradient). The corresponding macro-etched image is shown in Fig. \ref{fig:dynamic process control}(b).

We observe a significant improvement in the epitaxial growth regime, extending up to approximately 48 mm (zoomed in Fig. \ref{fig:dynamic process control}(c)), as indicated by the uniform color across the build, except for a thin streak of misoriented grains near the right edge. Optical micrographs (Fig. \ref{fig:dynamic process control}(d)) and the corresponding inverse pole figure map parallel to the build direction (Fig. \ref{fig:dynamic process control}(e)) further confirm the presence of an enlarged single-crystalline region, consistent with the macro-etched observations. However, beyond a build height of approximately 55 mm, stray grains nucleate and continue to grow toward the final build height.

\subsection{Conditions leading to stray grain formation}
\begin{figure}[!htb]%
\centering
\includegraphics[width=0.6\linewidth]{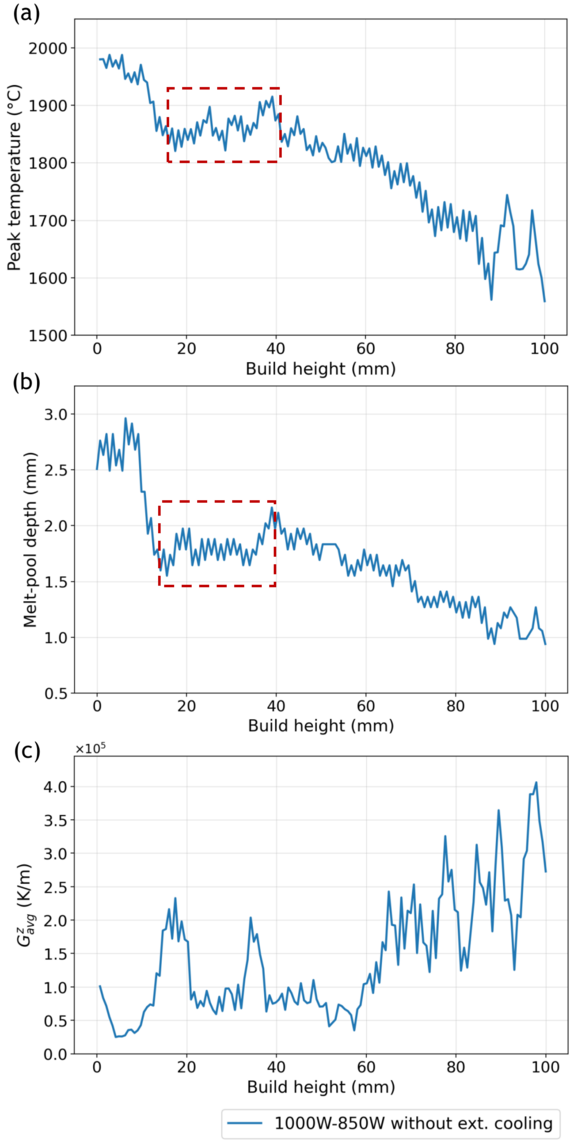}
\caption{Plots extracted from IR data for deposition with constant scan speed of 1400 mm/min and varying laser power as: 1200 W-1000 W reduction over 15 mm (9.52 W/layer), constant 1000 W up to 40 mm, remaining build height with 1000 W-850 W reduction in power (1.76 W/layer). (a) Peak melt-pool temperature, (b) melt-pool depth, and (c) average thermal gradient are all plotted as a function of build height. Red-dashed box in (a) and (b) highlights the region where the peak temperature and melt-pool depth increase.}\label{fig:peakt_d_grad_1000W-850W}
\end{figure}

Stray grain formation in both samples (Fig. \ref{fig:dynamic process control}(a,b)) can be understood by examining the melt-pool characteristics and average thermal gradients as a function of build height. Figure \ref{fig:peakt_d_grad_1000W-850W}(a) shows the evolution of peak melt-pool temperature with build height. 
The peak temperature is highest at the beginning of the build (approximately 1960 \textdegree C) and gradually decreases to about 1850 \textdegree C within the first 15 mm, consistent with the reduction in laser power from 1200 W to 1000 W.
From 15 mm, the peak temperature increases gradually, reaching a maximum around 40 mm (Fig. \ref{fig:peakt_d_grad_1000W-850W}(a)-red-dashed box), before the laser power reduction causes a decrease in peak temperature. The melt-pool depth follows a similar trend, as shown in Fig. \ref{fig:peakt_d_grad_1000W-850W}(b), with a drop to 1.5 mm at about 15--18 mm build height coinciding with the reduction in laser power. From here to 40 mm, there is a slight increase in melt-pool depth (Fig. \ref{fig:peakt_d_grad_1000W-850W}(b)-red-dashed box) before the magnitude begins to decrease as laser power is reduced. The average thermal gradients remain nearly the same until a build height of 50 mm, after which the magnitude starts to increase as the laser power decreases. The location for the onset of stray grain nucleation observed in the macro-etched image (Fig. \ref{fig:dynamic process control}(b)) matches the location where the melt-pool depth starts to decrease around 50--60 mm build height. These results indicate that, despite the increase in average thermal gradients as shown in Fig. \ref{fig:peakt_d_grad_1000W-850W}(c), the reduction in peak temperature and, consequently, melt-pool depth leads to insufficient remelting of stray grains in the previously deposited layers, thereby hindering sustained epitaxial growth. Thus, reducing laser power alone cannot simultaneously maintain adequate conditions for epitaxial growth, and retaining single-crystalline growth to greater heights requires control of heat accumulation, as well as steady-state peak temperatures and melt-pool depths. We explore one way to achieve it using an external cooling strategy. While some previous works \cite{chen2019, nie2021substratecooling} have incorporated substrate cooling, it is viable only for repairing or building small parts. For printing continuously long rod structures up to 100 mm, substrate cooling is ineffective, as heat extraction must be stronger and localized near the melt pool. This is only possible if the cooling setup moves with the laser head, so that the regions near the melt pool continue to be cooled at the same intensity. 

\section{External cooling for enhanced epitaxial growth}
We first conduct process simulations to understand the effect of external cooling on solidification conditions by varying the convective heat transfer coefficient. In the absence of the exact value of the convective heat transfer coefficient corresponding to rod printing experiments, we take the value $h_{conv}=270$ $W/m^2.K$ from our previous work \cite{bhure2025acta} as the starting point.

\subsection{Simulation setup} 
\label{sec:simulation_setup}
We use the process model described in our previous work \cite{bhure2025acta}, with the fluid-flow terms off, to quickly capture the trends in variation of melt-pool and solidification conditions for an 8 mm-diameter rod to a reasonable height of about 22 mm, using the same process parameters and scan strategy as in experiments.
We consider a simulation box of size $9.6\times 9.6\times 44$ $mm^3$ and initialize the bottom 4 mm height as metal ($\gamma=1$) and the remaining 40 mm height as gas ($\gamma=0$). The thermophysical properties and simulation parameters are the same as those for CMSX-4 alloy given in our previous work \cite{bhure2025acta}. The effective powder catchment efficiency is set to 33\% to match the 0.7 mm layer thickness maintained in the previous section. We initialize the metal region in the simulation at 1000 \textdegree C, representing the preheat condition prior to deposition.

\begin{figure}[!htb]%
    \centering
    \includegraphics[width=0.6\linewidth]{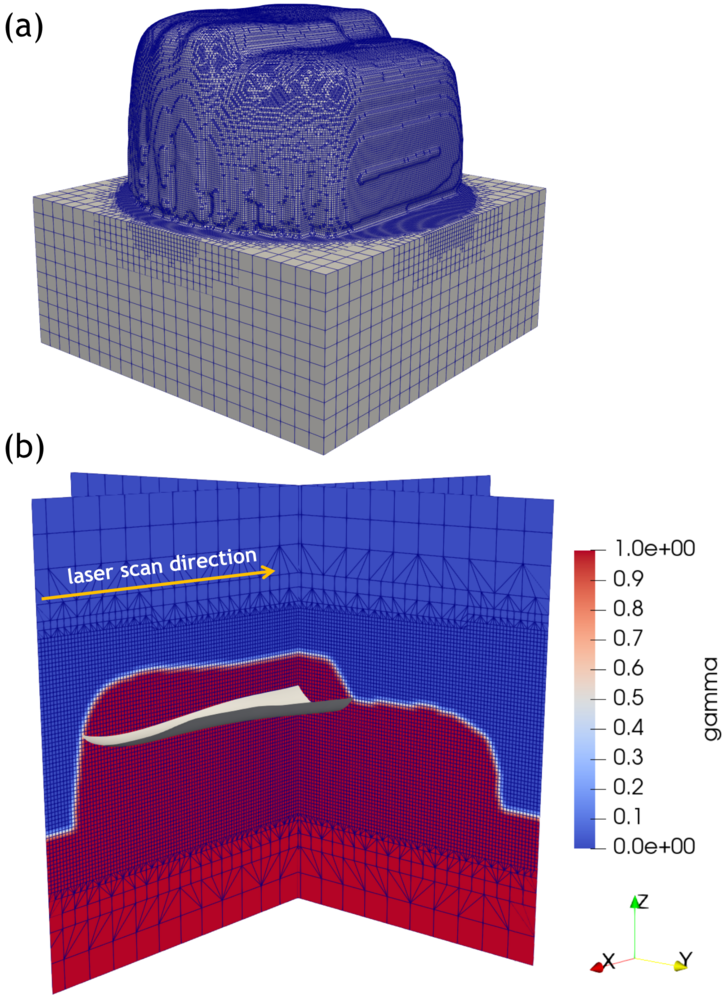}
    \caption{Simulation images showing (a) mesh refinement over the metal region of the rod, and (b) 2D cross-sections of the mesh showing three levels of mesh refinement (with the finest mesh near the metal-gas interface ($0<\gamma<1$)), and a 3D melt-pool contour in cream color.} 
    \label{fig:dynamic_mesh-refinement}
\end{figure}

To speed up the computation, we use the dynamic mesh-refinement feature in OpenFOAM-8, where we use three levels of mesh refinement, with the cell size being $5\times 10^{-5}$ m in the finest region, the same as the mesh size optimized in the previous work, at the metal-gas interface ($0<\gamma<1$), and throughout the whole melt-pool region to accurately capture the laser heat input, heat transfer within and across the melt-pool, latent heat evolution, and material deposition. The mesh coarsens gradually below the solidus temperature to a maximum cell size of $4\times 10^{-4}$. While choosing the refinement region, we make sure that the mesh stays refined far enough from the metal-gas interface and melt-pool at any given point within a layer, as shown in Fig. \ref{fig:dynamic_mesh-refinement}.

\subsection{Simulation discussion}
\label{sec:simulation}
\begin{figure}[!htb]%
    \centering
    \includegraphics[width=0.7\linewidth]{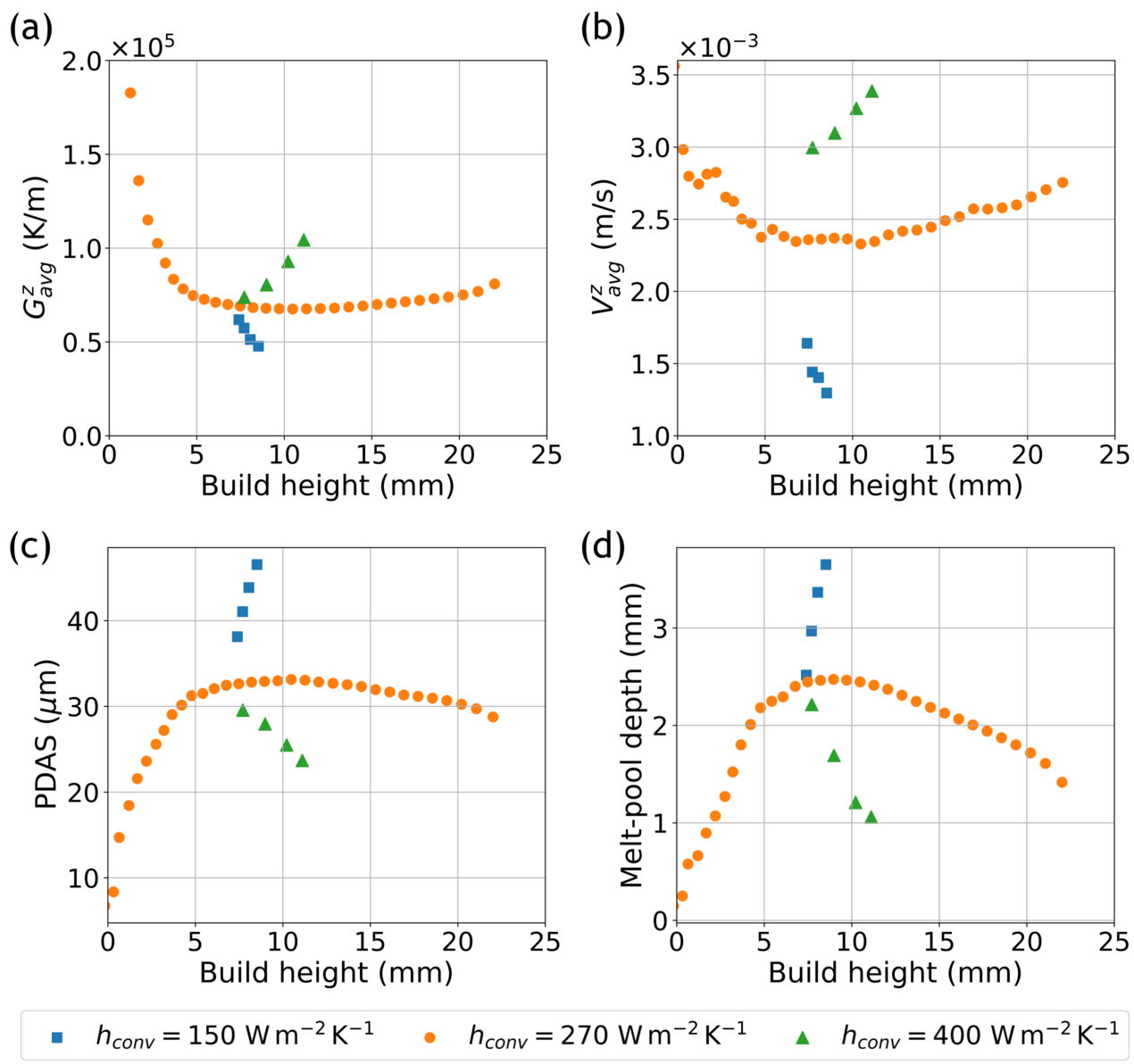}
    \caption{Plots obtained from process simulation showing the effect of external cooling by varying convective heat transfer coefficient ($h_{conv}$) on solidification behavior: (a) average thermal gradients ($G_{avg}^z$) and (b) solidification velocities ($V_{avg}^z$), both evaluated along [001] crystallographic growth direction, (c) primary dendrite arm spacing (PDAS), and (d) melt-pool depth, all plotted as a function of build height.} 
    \label{fig:sim_G_V_PDAS}
\end{figure}

We perform the process simulation with a convective heat transfer coefficient of $h_{conv}=270$ $W/m^2.K$ following the constant scan speed and variation in laser power same as that described in the previous section i.e. starting with the laser power of 1200 W and gradually reducing it to 1000 W at 9.52 W/layer up to a build height of 15 mm, and keeping the laser power constant thereafter. To understand the effect of this parameter on solidification and microstructure, we extract the melt-pool depth and the average thermal gradients ($G_{avg}^z$) on the melt-pool contour along [001] crystallographic growth direction using Paraview, an open-source data analysis and visualization software. We import this data along with melt-pool contours to compute average solidification velocity ($V_{avg}^z$) along [001] crystallographic growth direction using the procedure described in our previous work \cite{nalajala2026Materialia} based on Rappaz et al., and Anderson et al. \cite{rappaz1989development, anderson2010origin}. We then compute primary dendrite arm spacing (PDAS) using the relation derived for the CMSX-4 alloy from phase-field simulations in our previous work \cite{bhure2025acta}. The variation of all these parameters as a function of build height is shown in Fig. \ref{fig:sim_G_V_PDAS}.
We observe that for the first 5 mm of the build height, the solidification conditions are transient, with $G_{avg}^z$ reducing by about 2.75 times, which is explained by heating-dominated conditions, as evidenced by the melt-pool depth increasing to 2.15 mm at a build height of 5 mm. Analogously, $V_{avg}^z$ also reduces by $20\%$ because as the heat builds up, the melt-pools become larger and flatter, resulting in a reduction in solidification velocity. As a consequence of these conditions, PDAS increases to about 31 $\mu$m at 5 mm. Beyond this height, the variation in $G_{avg}^z$, $V_{avg}^z$, melt-pool depth, and thereby the PDAS is minimal as the system reaches a near-steady-state thermal condition.

Using this as a reference, we perform a couple of additional simulations by restarting the previous simulation at a build height of 7 mm with convective heat transfer coefficients of $h_{conv} = 150$ and 400 $W/m^2.K$, respectively.
Adopting the same post-processing procedure described above for these two cases, we plot them along with the previous result as shown in Fig. \ref{fig:sim_G_V_PDAS} with blue-square and green-triangle markers. We observe that $G_{avg}^z$ and $V_{avg}^z$ increase with higher convective cooling and decrease with lower convective cooling rate, whereas the melt-pool depth and PDAS decrease with higher cooling rates and increase with lower cooling rates.

Comparing the three conditions of convective heat transfer coefficient, we infer that for the imposed changes in laser power and for the build conditions with laser power 1000 W, there exists a near-steady-state condition for the effective heat-transfer coefficient close to $h_{conv}=270$ $W/m^2.K$. Any deviation will lead to heat accumulation or a drop in melt-pool temperatures. In the former case, the melt-pool depths will continue to increase, making the build infeasible, while in the latter case, the temperature gradients will increase at the expense of the reduction in melt-pool depths. Neither of these situations is conducive to continued epitaxial growth. Therefore, the simulations reveal the existence of a condition where there is a balance between the heat input and heat extraction, which provides motivation for engineering the cooling setup where steady-state solidification can be achieved. 

\subsection{Experimental optimization of external cooling parameters}
\label{external_cooling_exp}
\begin{figure}[!htb]%
    \centering
    \includegraphics[width=0.6\linewidth]{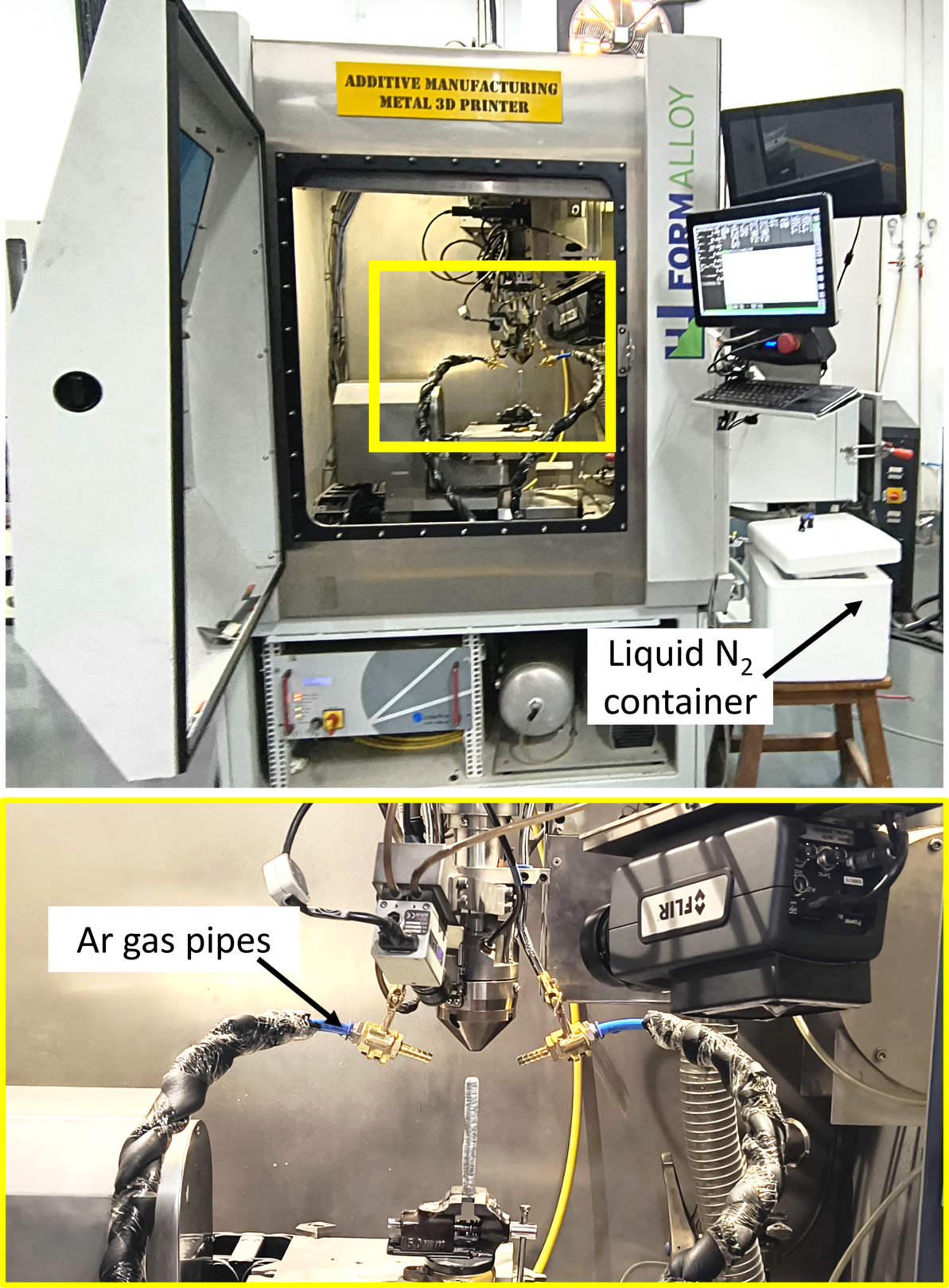}
    \caption{FormAlloy DED machine with external Argon cooling setup that moves synchronously along with the laser head.} \label{fig:cooling_setup}
\end{figure}

We design an external cooling setup using polyurethane (PU) tubing, through which argon gas flows after it has been cooled in a liquid-nitrogen container, as shown in Fig. \ref{fig:cooling_setup}.
The gas delivery lines are secured using clamps mounted on a fixture mechanically coupled to the laser head, allowing the cooling source to move synchronously with the deposition process. There are several handles available to control heat extraction, such as argon gas pressure, nozzle position, the location on the rod where the cooling nozzles are pointed, and the height at which we start cooling. For simplicity, we keep the argon gas pressure at 6 bar, the gas nozzles positioned approximately 38 mm away from the center of the laser nozzle (so as not to be influenced by the laser source), and the angle of the nozzles is chosen such that the gas flow does not influence the powder feed from the nozzle. With these conditions, gas cooling can be initiated at a height of around 20--30 mm, ensuring the cold gas impinges on the base of the build. 

\begin{figure}[!htb]%
    \centering
    \includegraphics[width=0.6\linewidth]{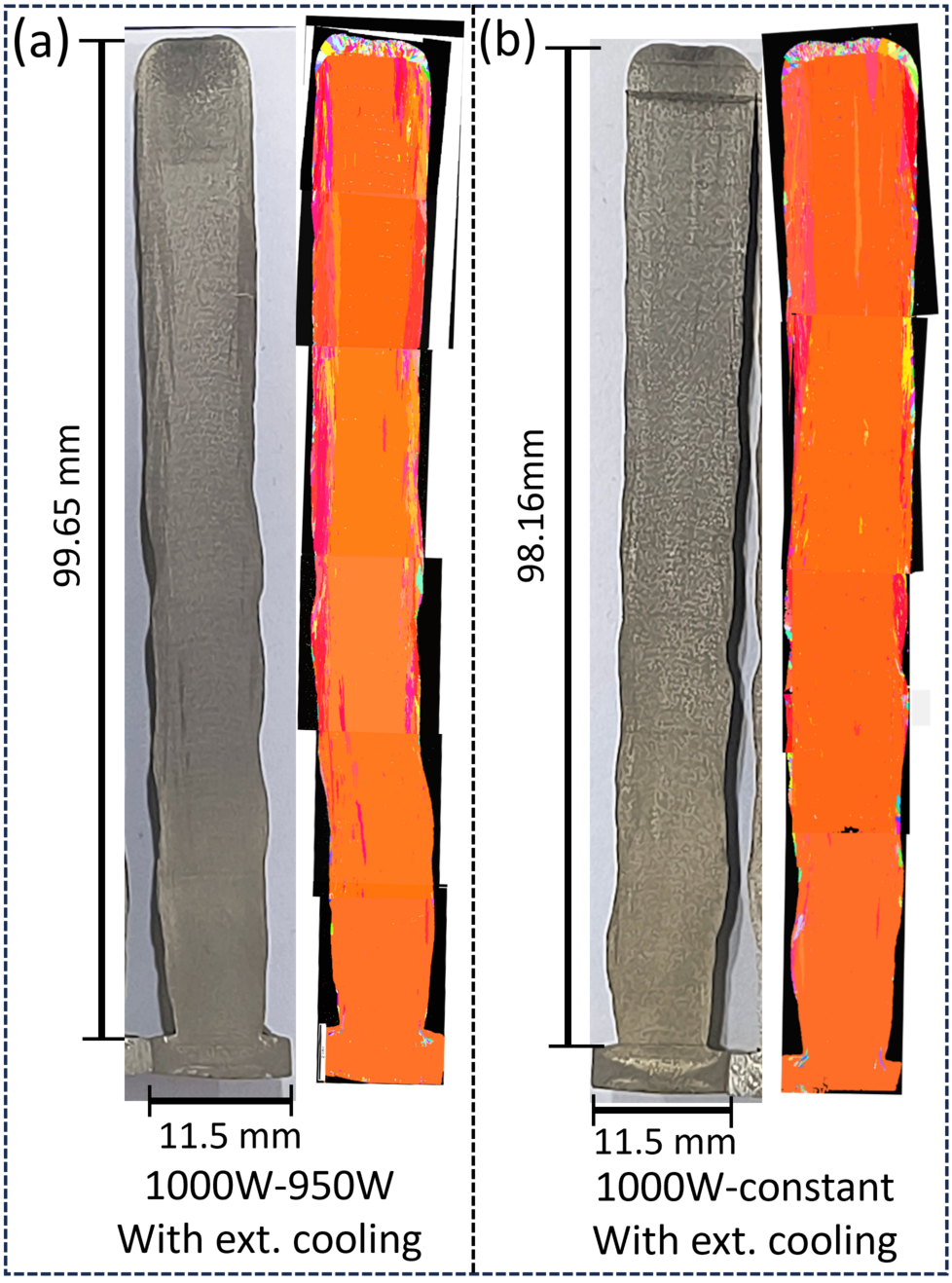}
    \caption{Macro-etched image and corresponding colored inverse pole figure maps parallel to the build direction for the samples printed with external cooling initiated at 22 mm build height with scan speed of 1400 mm/min and variation in laser power during printing. (a) Laser power reduced from 1200 W to 1000 W up to 15 mm, followed by a linear reduction from  1000 W to 950 W up to the final height, (b) laser power reduced from 1200 W to 1000 W up to 15 mm, and maintained 1000 W thereafter. Colored IPF maps are post-processed with a 10\textdegree grain average to remove noise from stitching of tens of scans.
    } \label{fig:final_rods}
\end{figure}

\begin{figure}[!htb]%
    \centering
    \includegraphics[width=0.58\linewidth]{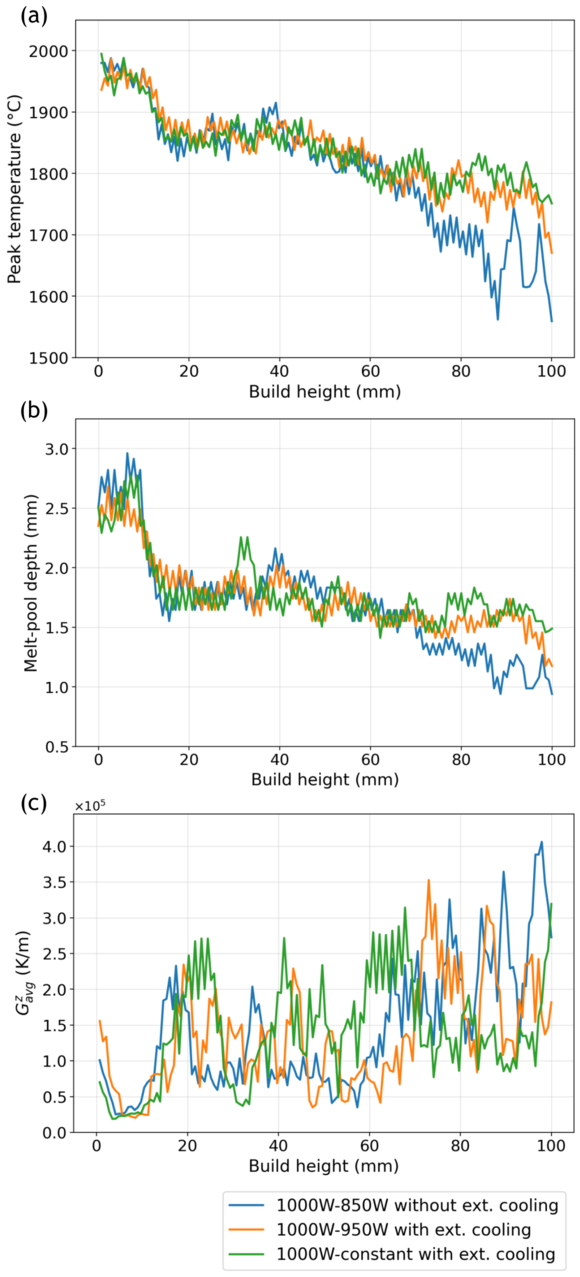}
    \caption{Plots extracted from IR data for deposition with constant scan speed of 1400 mm/min and varying laser power as: 1200 W-1000 W reduction over 15 mm (9.52 W/layer), constant 1000 W up to 40 mm, and remaining build height with variations listed in the legend. (a) Peak melt-pool temperature, (b) melt-pool depth, and (c) average thermal gradient are all plotted as a function of build height.
    } \label{fig:final_graphs}
\end{figure}

Initiating cooling at a build height of 22 mm, we find that near steady-state conditions are achieved. We conduct a trial by keeping the parameters the same up to 40 mm, then reducing the laser power to 950 W at the end height instead of 850 W. External cooling helps control heat accumulation, resulting in a build with good topology beyond 40 mm which was not possible before without cooling at laser powers higher than 850 W. Fig. \ref{fig:final_rods} (a) shows the macro-etched image and the corresponding colored inverse pole figure map parallel to the build direction for this parameter. 
The macro-etched image shows clean epitaxial growth from the bottom to the top at the center, with very few stray grains at both edges as we move towards the top of the build. The colored IPF map along the build direction shows a large single-crystalline region with [001] crystallographic orientation. Additionally, it is worth noting that the EBSD maps match the grain structure depicted in the macro-etched images for
this parameter. Significant improvement in the epitaxial growth regime can be explained by looking at the melt-pool characteristics shown in Fig. \ref{fig:final_graphs} with orange lines. 
For comparison, a trial conducted without cooling is also shown in blue. The peak temperature profile showed a trend similar to that of the no-cooling condition up to an approximate build height of 60 mm. Later on, an almost constant melt-pool temperature is maintained, in contrast to the no-cooling condition, where the temperature drops significantly due to reduced laser power. Melt-pool depth is also stable and higher than in the no-cooling condition. Consistently higher melt-pool temperature, thereby melt-pool depth, in addition to the higher average thermal gradients, resulted in significant improvement in the epitaxial growth regime. 

To examine whether a further increase in power is possible to improve epitaxial growth, we conduct another trial in which the laser power is held constant at 1000 W from 15 mm to the top. The macro-etched image and corresponding colored inverse pole figure map can be seen in Fig. \ref{fig:final_rods}(b). Compared with the previous sample, there is a slight improvement in the epitaxial growth regime. This is possibly attributed to the slightly increased melt-pool temperature and depth, as shown in Fig. \ref{fig:final_graphs} by the green lines. These observations confirm that, in addition to maintaining higher average thermal gradients, maintaining a consistent melt-pool peak temperature and stable melt-pool depth helps maximize epitaxial growth. Based on these observations, we select the 1000 W constant condition for further study, in which we analyze the solidification microstructure and conduct additional heat-treatment optimization trials on the sample printed with these parameters.

\section{Solidification microstructure}
\begin{figure}[!htb]%
    \centering
    \includegraphics[width=0.6\linewidth]{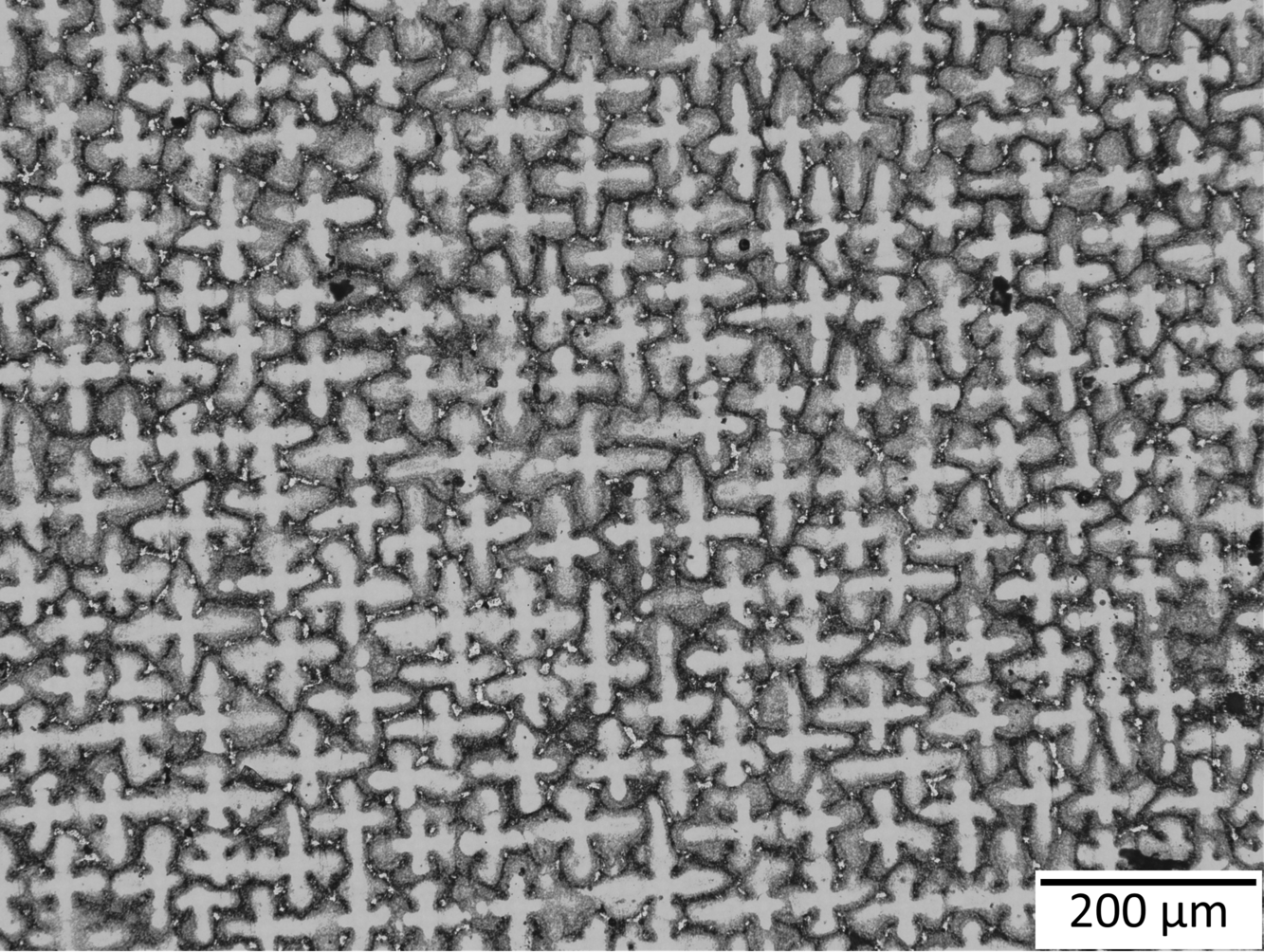}
    \caption{Transverse optical micrograph acquired at a build height of 45 mm, printed with external gas cooling at a scan speed of 1400 mm/min, with laser power reduced from 1200 W to 1000 W up to 15 mm, and maintained at 1000 W thereafter.
    } \label{fig:transverse}
\end{figure}

\begin{figure}[!htb]%
    \includegraphics[width=0.7\linewidth]{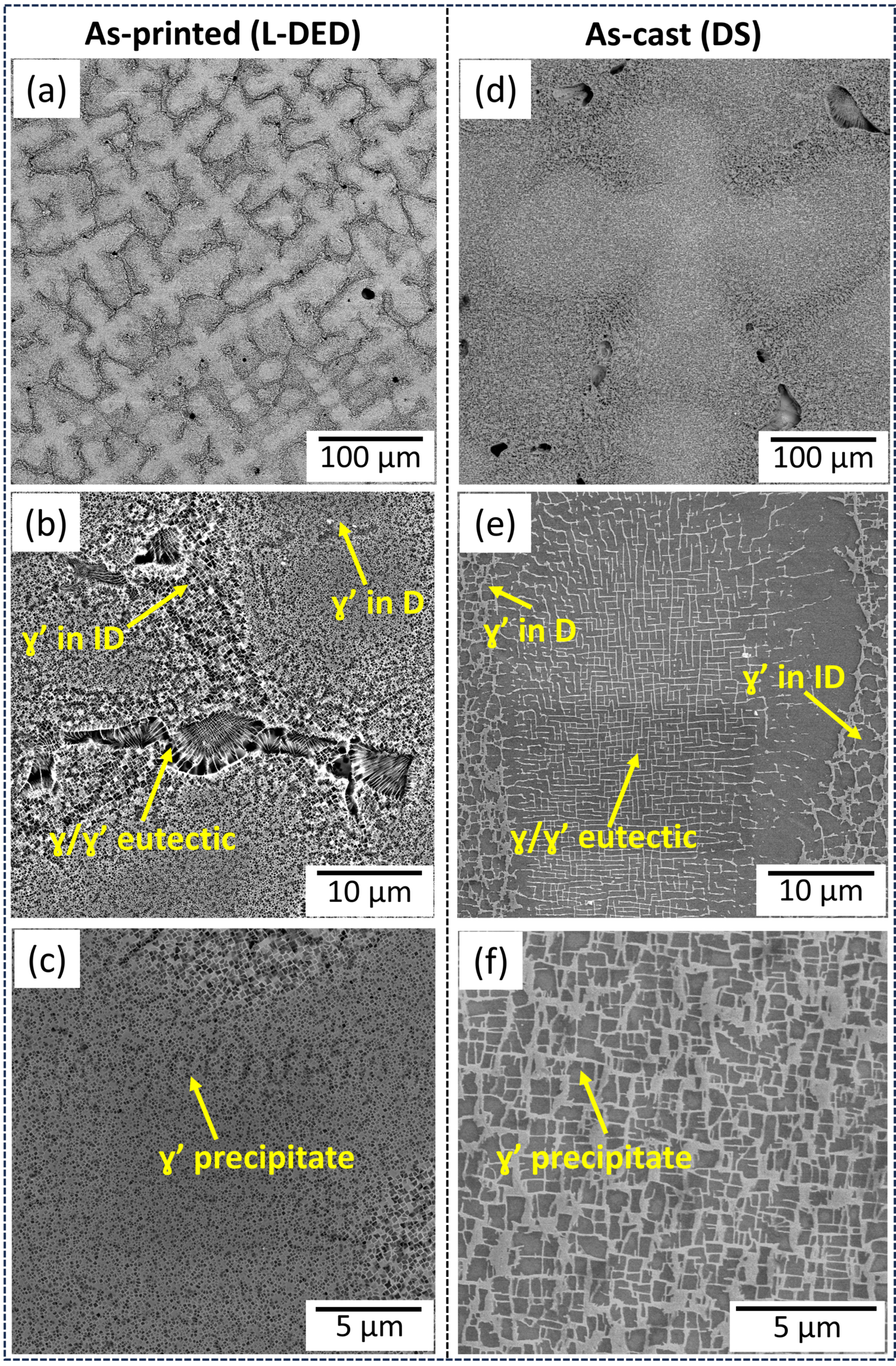}
    \centering
    \caption{(a-c) SEM images of CMSX-4 processed by L-DED, and (d-f) by DS showing the dendritic microstructure, inter-dendritic $\gamma$/$\gamma\prime
    $ eutectic and $\gamma\prime
    $ precipitate distribution in the matrix, respectively, in the as-cast conditions. D and ID denote the dendritic and inter-dendritic regions, respectively.
    }\label{fig:sem_as_printed}
\end{figure}

We perform microstructural analysis at various locations across the complete build height for the sample printed with external cooling using a constant 1000 W power (Fig. \ref{fig:final_rods}(b)). A representative transverse optical micrograph at a build height of 45 mm is shown in Fig. \ref{fig:transverse}, revealing a single dendrite orientation over a larger area. 
Fig. \ref{fig:sem_as_printed}(a-c) shows SEM images at the same location, revealing the presence of a $\gamma$/$\gamma\prime$ eutectic in the inter-dendritic regions and an inhomogeneous size distribution of $\gamma\prime$ precipitates in the dendritic and inter-dendritic regions, respectively. 
For comparison, in-house processed CMSX-4 single-crystals produced by directional solidification (DS) are also shown in Fig. \ref{fig:sem_as_printed}(d-f), revealing a coarser dendritic microstructure with a PDAS of 300 $\mu$m, with a larger eutectic length scale and coarser $\gamma\prime$ precipitates compared with the L-DED processed sample. To understand elemental segregation behavior in L-DED-processed samples, EPMA point analyses are performed in the dendritic and interdendritic regions. Table \ref{tab:comp_analysis} summarizes the measured segregation coefficient values ($k=c_s/c_l$) for all the alloying elements present in the CMSX-4 alloy. Al, Ti, Ta, and Mo, being $k<1$, segregate to the inter-dendritic regions, while W and Re, $k>1$, are concentrated at the core of the dendrite. 

\begin{figure}[!htb]%
    \centering
    \includegraphics[width=0.7\linewidth]{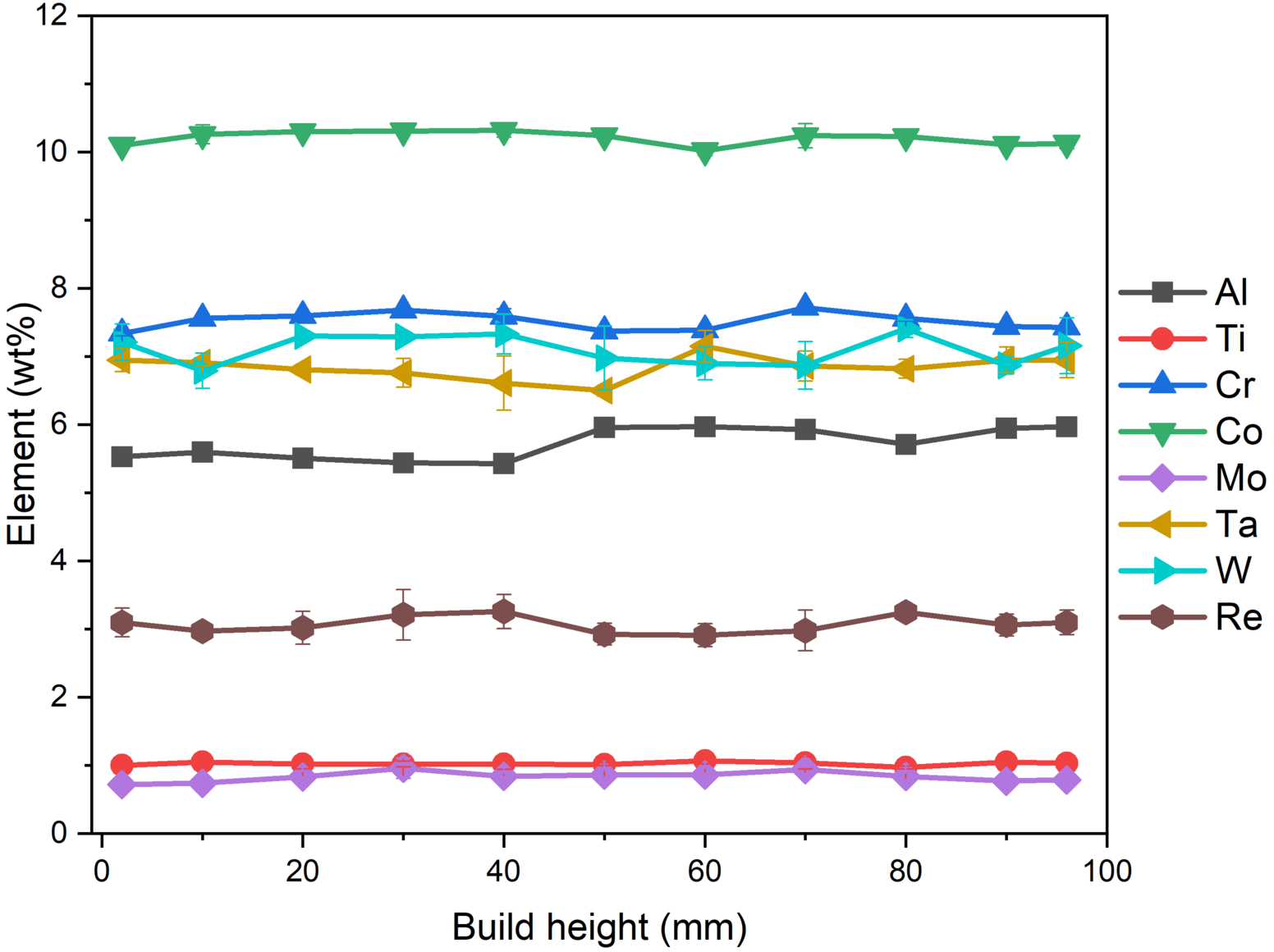}
    \caption{Variation of the chemical composition of the individual alloying element with build height for the sample printed with external gas cooling at a scan speed of 1400 mm/min, with laser power reduced from 1200 W to 1000 W up to 15 mm, and maintained at 1000 W thereafter.
    } \label{fig:composition}
\end{figure}

\begin{figure}[!htb]%
    \centering
    \includegraphics[width=0.7\linewidth]{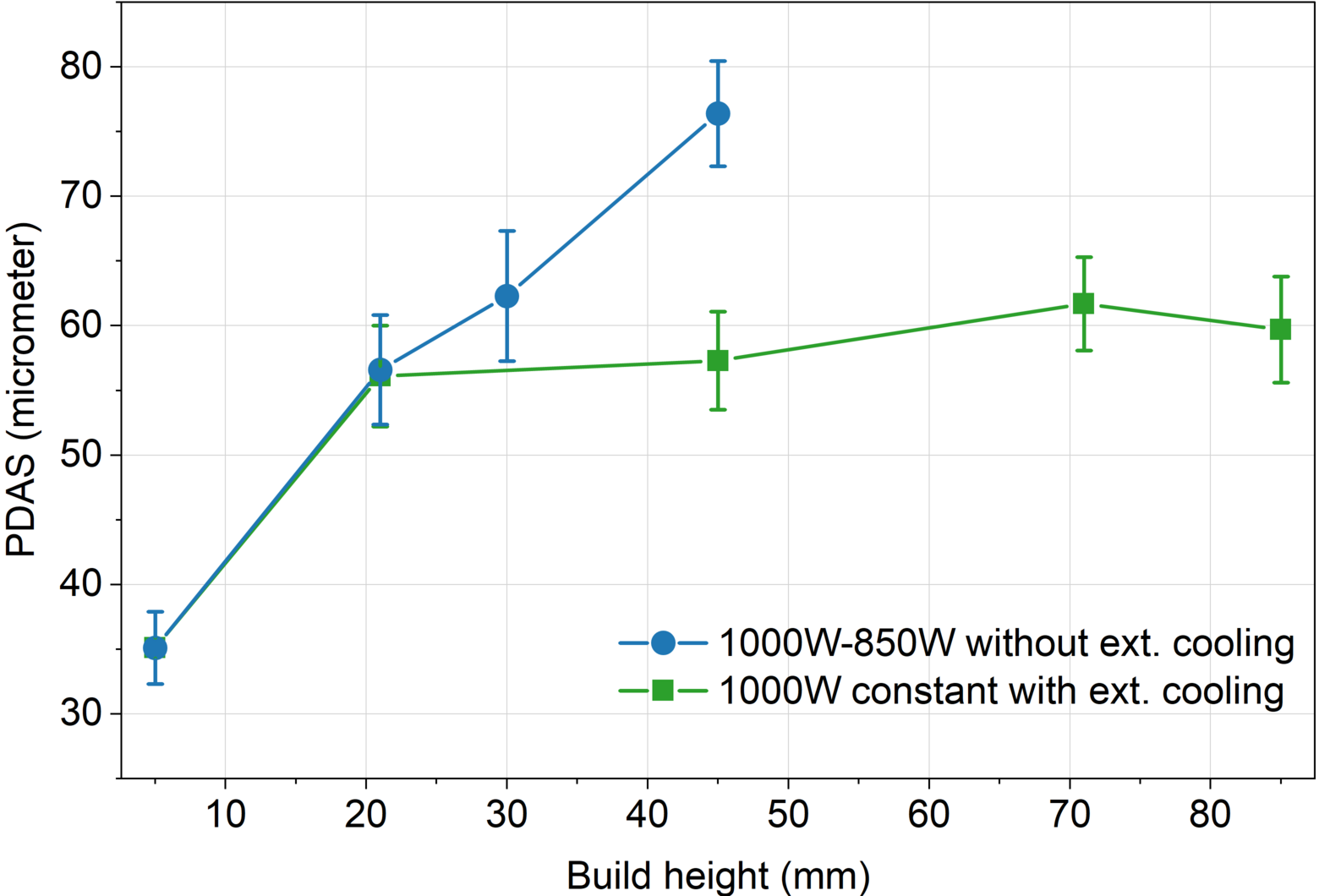}
    \caption{Variation of PDAS with build height for the samples printed at a scan speed of 1400 mm/min; without external cooling, with laser power reduced from 1000 W to 850 W starting at 40 mm build height, and with external cooling, at a constant 1000 W throughout the build.
    } \label{fig:pdas}
\end{figure}

\begin{table}[!hbt]
    \caption{Elemental composition (wt\%) and corresponding segregation coefficient (k) values in as-printed condition at 45 mm build height for L-DED processed sample.}
    \label{tab:comp_analysis}
    \centering
    \begin{tabular}{llll}
        \toprule
        Elements & Dendrite  & Inter-dendrite & k\\
        \midrule
        W & 8.6 $\pm$ 0.14& 5.74 $\pm$ 0.76& 1.52 $\pm$ 0.2\\
        Co & 11.45 $\pm$ 0.08& 11.32 $\pm$ 0.66 & 1.01 $\pm$ 0.07\\ 
        Ta & 3.01 $\pm$ 0.07& 4.66 $\pm$ 1.22& 0.69 $\pm$ 0.2\\
        Ti & 0.61 $\pm$ 0.03& 1.2 $\pm$ 0.35& 0.54$\pm$ 0.15\\
        Mo & 0.61 $\pm$ 0.07& 0.84 $\pm$ 0.18& 0.76 $\pm$ 0.18\\
        Re & 3.93 $\pm$ 0.14& 2.83 $\pm$ 0.56& 1.44 $\pm$ 0.35\\
        Cr & 7.42 $\pm$ 0.11& 9.54 $\pm$ 1.1& 0.77 $\pm$ 0.1\\
        Al & 4.49 $\pm$ 0.08& 4.7 $\pm$ 0.44& 0.96 $\pm$ 0.11\\
        Ni & 61.03 $\pm$ 0.44& 58.91 $\pm$ 1.92& 1.04 $\pm$ 0.04\\
        \bottomrule
    \end{tabular}
\end{table}
To characterize the macrosegregation, the variation in the chemical composition of each element is measured along the build height and summarized in Fig. \ref{fig:composition}.
The chemical composition of all elements remains relatively unchanged, indicating the absence of macrosegregation of alloying elements. In addition, PDAS measured from transverse sections at build heights of 5, 21, 45, and 71 mm is summarized in Fig. \ref{fig:pdas} with a green line. 
The PDAS is finest near the base of the build, measuring approximately 35.1 $\mu$m at 5 mm, and increases to about 56.1 $\mu$m at 21 mm. Beyond this height, the PDAS remained relatively constant, indicating that the solidification conditions become relatively stable with increasing build height, which is in line with the melt pool characteristics discussed in the previous section \ref{external_cooling_exp} (Fig. \ref{fig:final_graphs}, green-colored lines). In contrast, the sample printed without external cooling exhibits a continuous increase in PDAS, as shown in Fig. \ref{fig:pdas} by the blue-colored line, consistent with progressive heat accumulation. These observations are consistent with the simulations performed with $h_{conv}=270$ $W/m^2.K$ and $h_{conv}=150$ $W/m^2.K$, respectively, in Subsection \ref{sec:simulation}. 

Overall, the results indicate relatively uniform chemical composition and solidification behavior with build height. However, significant microstructural and compositional heterogeneities persist at the microscale, including eutectics, an inhomogeneous distribution of $\gamma\prime$ precipitates, and microsegregation of various alloying elements. These heterogeneities can limit the high-temperature properties of DED-processed samples. Therefore, an appropriate heat-treatment is required to reduce the chemical and microstructural heterogeneities associated with the as-printed microstructure. 

\section{Heat-treatment optimization}
To begin, we adopt the standard heat-treatment cycle used for conventionally grown CMSX-4 single-crystals as the reference \cite{Matache02092016, rzyankina2016, wilson2003effect}, which consists of a multi-step solutionizing treatment followed by two-step aging, as summarized in Table \ref{tab:heat_treatment_DS}. 
\begin{table}[!htb]
    \caption{Heat-treatment cycle followed in this study; heating rate 3 \textdegree C/min.}
    \centering
    \begin{tabular}{p{0.15\textwidth}p{0.40\textwidth}p{0.15\textwidth}p{0.15\textwidth}}
        \toprule
        \rule{0pt}{2.5ex}
        Sample no. & Solutionizing  & Aging1 & Aging2\\
        \midrule
        \rule{0pt}{2.5ex}
        I (L-DED)  & 1277 \textdegree C/4 h, 1287 \textdegree C/2 h, 1296 \textdegree C/3 h, 1304 \textdegree C/3 h, 1313 \textdegree C/2 h, 1316 \textdegree C/5 h & 1140 \textdegree C/6 h & 871 \textdegree C/20 h \\\\
        II (DS)& 1277 \textdegree C/4 h, 1287 \textdegree C/2 h, 1296 \textdegree C/3 h, 1304 \textdegree C/3 h, 1313 \textdegree C/2 h, 1316 \textdegree C/5 h  & 1140 \textdegree C/6 h & 871 \textdegree C/20 h \\\\
        III (L-DED) & 1304 \textdegree C/1 h &-&-\\\\
        IV (L-DED)&1304 \textdegree C/3 h, 1313 \textdegree C/2 h, 1316 \textdegree C/5 h & 1140 \textdegree C/6 h&871 \textdegree C/20 h\\\\
        V (DS)&1304 \textdegree C/3 h, 1313 \textdegree C/2 h, 1316 \textdegree C/5 h &- & -\\
        \bottomrule
        \multicolumn{4}{l}{\rule{0pt}{2.5ex}\footnotesize L-DED: laser-directed energy deposition; DS: directional solidification}
        \label{tab:heat_treatment_DS}
    \end{tabular}
\end{table}

\begin{figure}[!htb]%
    \includegraphics[width=0.7\linewidth]{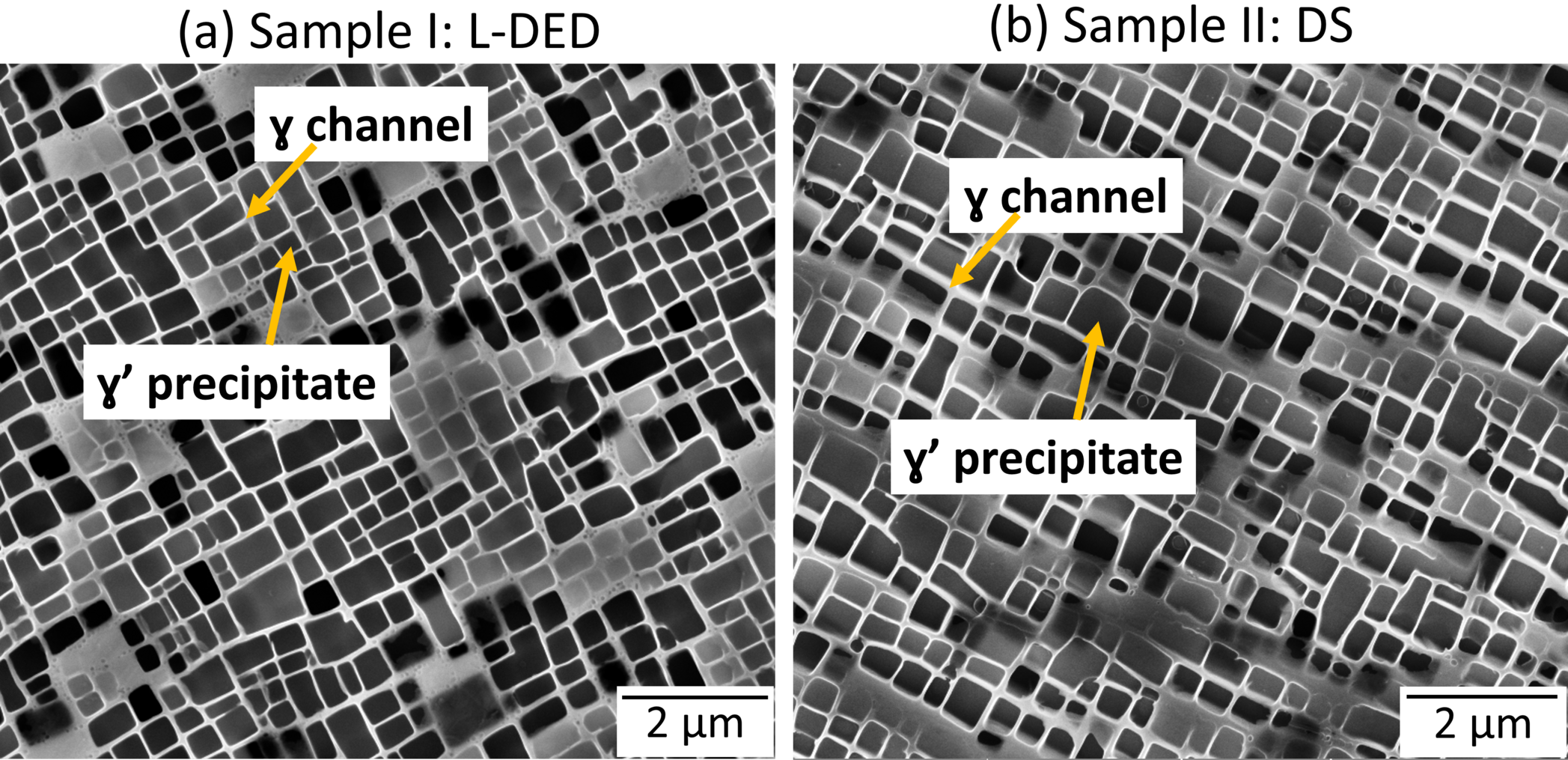}
    \centering
    \caption{SEM images revealing $\gamma\prime$ precipitates after double aging produced by (a) L-DED and (b) directional solidification using a vertical Bridgman furnace.
    } \label{fig:ds_am_comparison}
\end{figure}

Microstructural analysis after solutionizing confirms near-complete dissolution of the inter-dendritic $\gamma$/$\gamma\prime$ eutectic in the matrix, while subsequent aging results in homogeneous cuboidal $\gamma\prime$ precipitates as shown in Fig. \ref{fig:ds_am_comparison}(a). 
For comparison, we apply the same heat-treatment cycle to in-house-grown CMSX-4 single-crystals produced by directional solidification (DS) in a vertical Bridgman furnace, as shown in Fig. \ref{fig:ds_am_comparison}(b). Both samples reveal similar characteristics of $\gamma\prime$ after heat-treatment, with mean sizes of 420 $\pm$ 90 nm and 458 $\pm$ 80 nm, respectively. These observations are consistent with the microstructures reported for CMSX-4 single-crystals in the literature \cite{ramsperger2015solution, wilson2003effect}. The conventional solutionizing heat-treatment schedule is designed to account for a relatively coarser microstructural length scale after solidification, as shown in Fig. \ref{fig:sem_as_printed}(d-f), and for the associated severe microsegregation in conventionally grown CMSX-4. Severe microsegregation results in local compositional variations that can lower the local melting temperature in the inter-dendritic region close to eutectic temperature, thereby increasing the risk of incipient melting during high-temperature solutionizing \cite{wilson2003effect, WANG20143112, MIRAK2022112449}. In addition, the relatively larger PDAS results in longer diffusion distances for alloying elements during homogenization. The low diffusivity of the refractory elements such as Re and W further limits the homogenization kinetics \cite{karunaratne2000}. Therefore, directly applying a high solutionizing temperature can promote incipient melting of the eutectic phases before sufficient homogenization is achieved. To minimize this risk while achieving near-homogeneity of all alloying elements, the conventional solutionizing process begins at a relatively low temperature, 1277 \textdegree C, and then progressively increases to 1316 \textdegree C.

\begin{figure}[!htb]%
    \includegraphics[width=0.6\linewidth]{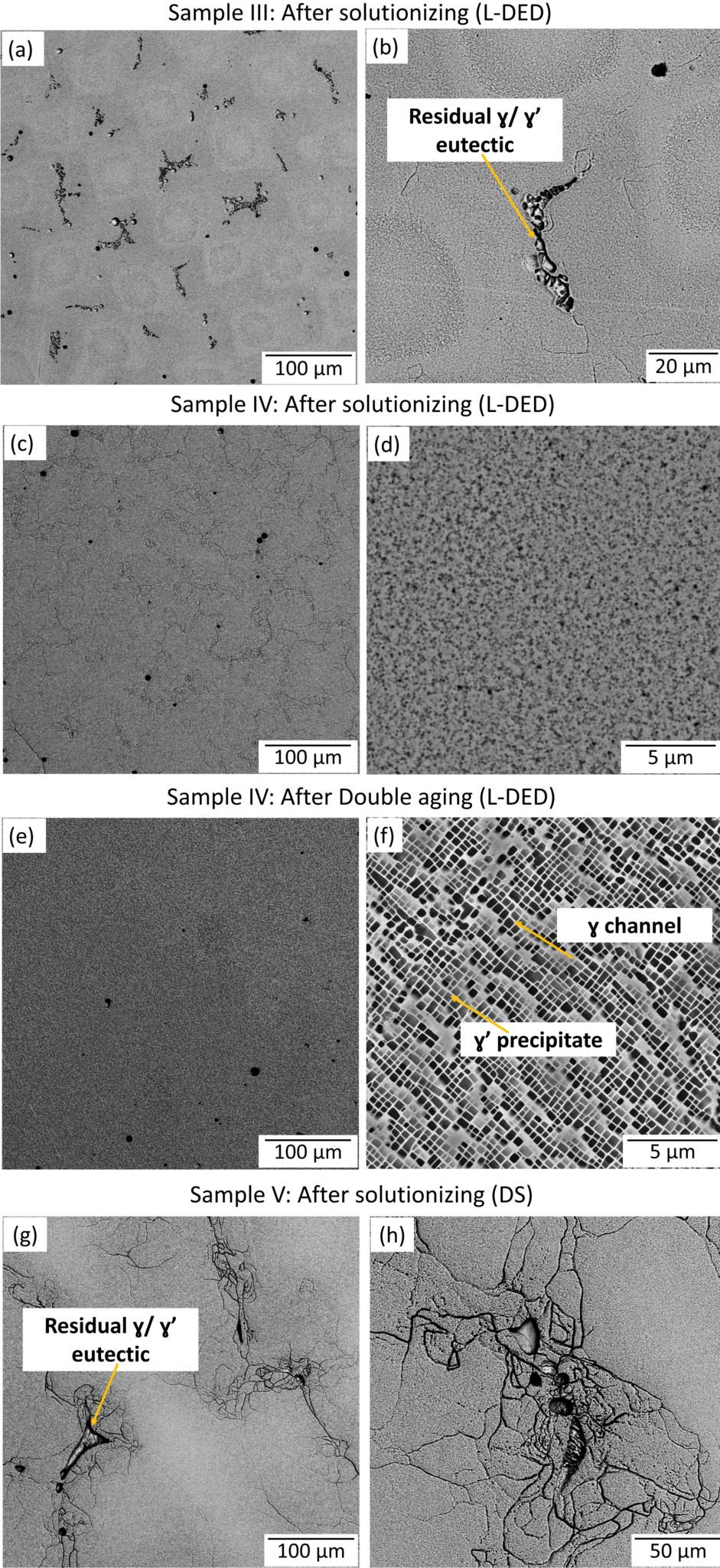}
    \centering
    \caption{(a,b) SEM images showing partial eutectic dissolution after solutionizing at 1304 \textdegree C/1 h at different magnifications; (c,d) showing complete eutectic dissolution after solutionizing at 1304 \textdegree C/3 h, 1313 \textdegree C/2 h and 1316 \textdegree C/5h; (e,f) reveal cuboidal $\gamma\prime$ precipitates after double aging for the L-DED processed samples at different magnifications and, (g,h) show partial eutectic dissolution after solutionizing at 1304 \textdegree C/3 h, 1313 \textdegree C/2 h, and 1316 \textdegree C/5 h for the DS-processed sample.
    } \label{fig:sem_heat_treated}
\end{figure}

\begin{figure}[!htb]%
    \includegraphics[width=0.8\linewidth]{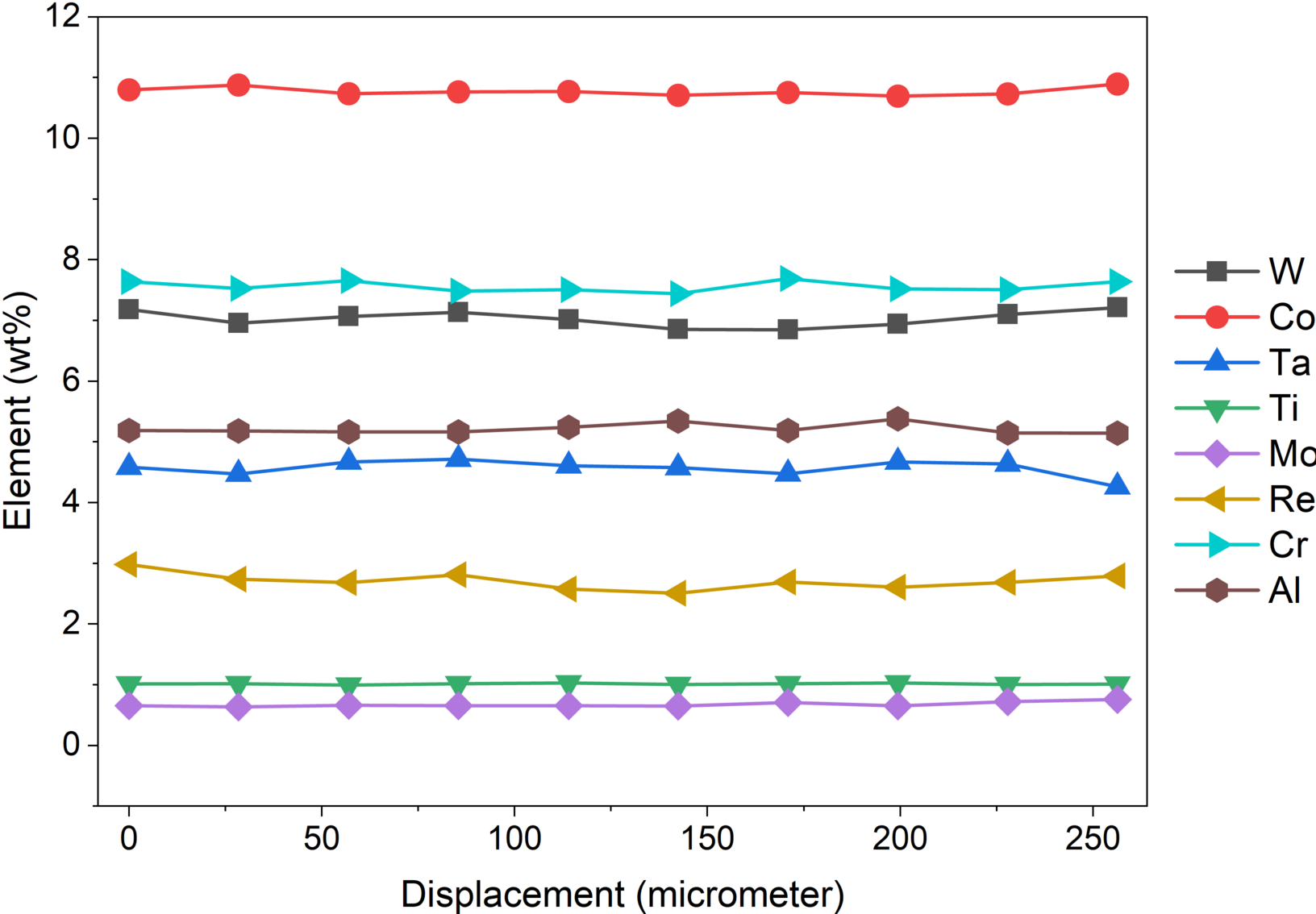}
    \centering
    \caption{A line scan showing the variation of the chemical composition of the individual alloying elements after solutionizing for the sample printed with external gas cooling at a scan speed of 1400 mm/min, with laser power reduced from 1200 W to 1000 W up to 15 mm, and maintained at 1000 W thereafter.
    } \label{fig:compo_heat_treated}
\end{figure}

\begin{figure}[!htb]%
    \includegraphics[width=0.99\linewidth]{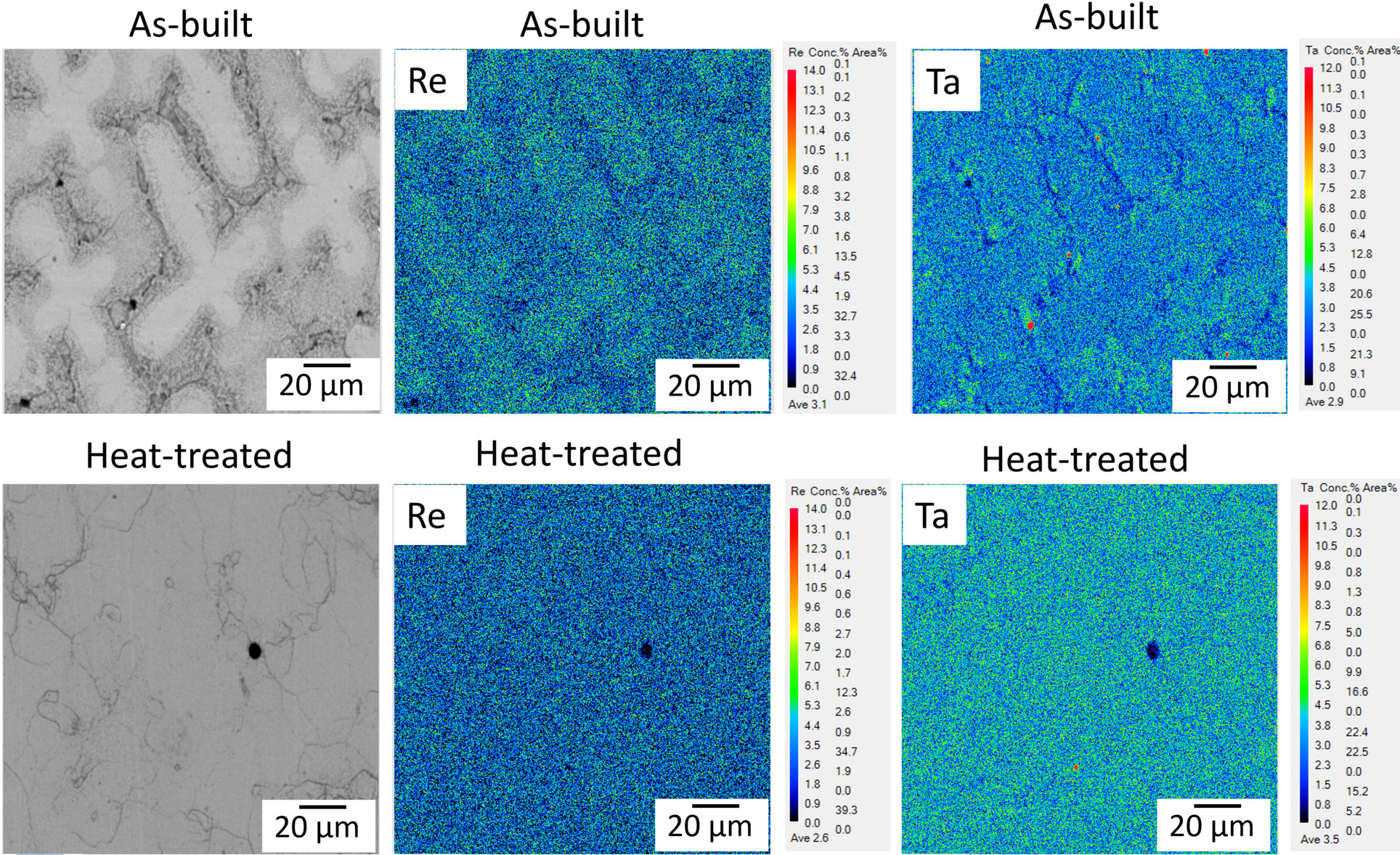}
    \centering
    \caption{Large area map reveals the distribution of Re and Ta in the as-printed and solutionized conditions, respectively, for the sample printed with external gas cooling at a scan speed of 1400 mm/min, with laser power reduced from 1200 W to 1000 W up to 15 mm, and maintained at 1000 W thereafter.
    } \label{fig:compo_map}
\end{figure}

The design of the heat-treatment schedule is therefore microstructure-specific. In contrast to the conventionally grown CMSX-4, the DED-processed single-crystals exhibit a significantly refined microstructural length scale approximately 4-8 times smaller, together with a shorter eutectic length scale \cite{ramsperger2016selective, WANG20143112, WAGNER2004270, LOPEZGALILEA2019100874}. Therefore, a tailored heat-treatment schedule is required for the DED-processed samples. The reduced diffusion distances suggest that homogenization can be achieved without the initial low-temperature stages used for conventionally grown samples. This is also consistent with the reports showing that SEBM-processed CMSX-4 single-crystals, with a PDAS of approximately 10 $\mu$m, require significantly shorter homogenization times \cite {ramsperger2015solution, ramsperger2016microstructure}. Based on this, we perform an initial trial at 1304 \textdegree C for 1 hour, followed by controlled quenching in Argon. Microstructural analysis reveals partial eutectic dissolution, indicating that the selected time is insufficient for complete eutectic dissolution at this temperature, as seen in Fig. \ref{fig:sem_heat_treated}(a,b) at different magnifications. 
Therefore, we retain the subsequent high-temperature segments in the conventional heat-treatment cycle (1304 \textdegree C/ 3 h, 1313 \textdegree C/ 2 h, 1316 \textdegree C/ 5 h). This modified solutionizing treatment results in complete eutectic dissolution and compositional homogenization across the DED-processed samples. Fig. \ref{fig:sem_heat_treated}(c, d) shows the homogenized microstructure after the solutionizing heat-treatment cycle, followed by double aging, revealing the presence of homogeneous cuboidal $ \gamma\prime$ (Fig. \ref{fig:sem_heat_treated} (e,f) with a volume fraction (69\%) and a mean size of 438 $\pm$ 82 nm, respectively at different magnifications. Further, to evaluate compositional uniformity, we perform a line scan over a distance of 260 $\mu$m, shown in Fig. \ref{fig:compo_heat_treated}. It reveals a near-homogenization of the chemical composition of all alloying elements.
Furthermore, to assess compositional uniformity over a large area, we obtain elemental maps of Re and Ta in the as-printed and solutionizing conditions as shown in Fig. \ref{fig:compo_map}. Re is selected because it preferentially segregates to the dendritic regions, whereas Ta preferentially segregates to the inter-dendritic regions. The elemental map shows a more homogeneous distribution of both Re and Ta after the solutionizing \cite{ramsperger2016microstructure}. These observations confirm that the optimized solutionizing treatment resulted in microstructural and compositional uniformity without affecting the aging response \cite{ramsperger2016selective}. 
Furthermore, we apply the same optimized heat-treatment cycle to conventionally grown CMSX-4 single-crystals processed using an in-house-developed vertical Bridgman furnace to assess the influence of the microstructural length scale on the required solutionizing. Microstructural analysis after the solutionizing reveals partial dissolution of the eutectic phases, as shown in Fig. \ref{fig:sem_heat_treated} (g, h) at different magnifications. 
This observation indicates that the solutionizing time required for the L-DED processed sample is shorter than that of the DS processed CMXS-4, owing to the refined microstructural length scale and possibly differences in the size and chemistry of the eutectic phases \cite{basak2017microstructures}. Consequently, the optimized heat-treatment schedule reduces the total solutionizing time by 47\% compared with the conventional heat-treatment cycle. This observation is consistent with the previous reports showing that the refined microstructural length scale in SEBM-processed CMSX-4 single-crystals enables shorter homogenization times \cite{ramsperger2016selective, LOPEZGALILEA2019100874}. The reduced heat-treatment cycle for DED-processed samples is specified in Table \ref{tab:heat_treatment_DED}.

\section{Conclusions}
In this paper, we demonstrate the feasibility of producing bulk single-crystalline microstructures via L-DED through a combination of experiments and simulations. Maintaining constant laser power and scan speed limits epitaxial growth to a certain height as the local thermal conditions change with the evolving thermal history, resulting in the loss of topology and promoting stray grain nucleation as thermal gradients decrease due to heat accumulation. Dynamic parameter variation helps extend epitaxial growth to an extent; however, beyond a certain build height, thermal conditions can no longer be maintained solely through parameter variation. Although reducing laser power helps control heat accumulation, excessive power reduction decreases the melt-pool temperature and depth, thereby limiting remelting of stray grains in previously deposited layers and hindering continued epitaxial growth. 
Furthermore, process simulations confirm the existence of a steady-state condition during printing in which heat input and heat extraction remain balanced. An appropriate design of an external cooling setup provides efficient heat extraction, allowing a higher constant deposition power to be maintained throughout the build, thereby sustaining the near-constant thermal conditions required to achieve extended epitaxial growth. The uniform solidification conditions are evidenced by near-constant melt-pool temperature, melt-pool depth, and PDAS across the build height. The resulting microstructure exhibits a finer PDAS than that of conventional processing, enabling optimization of the solutionizing heat-treatment cycle and a 47\% reduction in total solutionizing time. Subsequent aging reveals similar $\gamma\prime$ characteristics to those of conventionally grown CMSX-4 single-crystals. The results are promising for extending the use of L-DED in the single-crystal manufacturing of bulk samples, at the very least for simpler geometries.

\begin{table}[htbp]
    \centering
    \caption{Heat treament schedule implemented for DED processed CMSX-4 single crystals, heating rate 3\textdegree C, followed by argon purging at 2 bar pressure}
    \label{tab:heat_treatment_DED}
    \begin{tabular}{ccc}
        \toprule
        Solutionizing & Ageing 1 & Ageing 2 \\
        \midrule
        1304\,$^\circ$C/3\,h + 1313\,$^\circ$C/2\,h + 1316\,$^\circ$C/5\,h & 1140\,$^\circ$C/6\,h & 871\,$^\circ$C/20\,h \\
        \bottomrule
    \end{tabular}
\end{table}

\section{Acknowledgments} 
This research is supported by the ARDB grant and the DRDO DIA-RCoE grant, and the second author receives financial support through the PMRF scheme. We thank Prof. Dipankar Banerjee and Prof. Gunther Eggeler for providing the CMSX-4 single-crystal substrates, and the Smart Factory at CPDM, IISc, for granting us access to the FormAlloy L-DED machine, and Prashanth Soundappan and Siva S. for helping us print samples. We also acknowledge the AFMM, IISc, for providing access to their advanced characterization and microscopy facilities, namely the Thermo Fisher Helios G4-UX for EBSD and the JEOL JXA-8230 for EPMA.

\section*{Declarations}
The authors declare that they have no conflicts of interest.
All data related to paper can be provided upon request. Materials used for the work may be provided upon special request, but needs to be returned. The codes used for the simulations and analysis can be provided upon request. 

The contributions of the authors are as follows:

Divya Nalajala: Writing – review and editing, Writing – original draft, Visualization, Validation, Software, Methodology, Investigation, Formal analysis, Data curation, Conceptualization.

Swapnil Bhure: Writing – review and editing, Writing – original
draft, Visualization, Validation, Software, Methodology, Investigation,
Formal analysis, Data curation, Conceptualization.  

R. Ajithkannan: Writing – review and editing, Validation, Methodology, Investigation,
Formal analysis, Conceptualization.  

Shubhang Galagali: Writing – review and editing, Visualization, Validation, Software, Methodology, Investigation,
Formal analysis, Data curation.  

Abhik Choudhury: Writing – review and
editing, Writing – original draft, Visualization, Validation, Supervision,
Software, Resources, Project administration, Methodology, Investigation, Funding acquisition, Formal analysis, Data curation, Conceptualization.

\bibliography{manuscript}

\end{document}